\documentclass[structabstract]{aa}
\usepackage{graphicx}
\usepackage{txfonts}
\usepackage{natbib}
\newcommand{\methanol}{$\mbox{CH}_{3}\mbox{OH}$}
\newcommand{\hmol}{$\mbox{H}_{2}$}

\newcommand{\kms}{km\,s$^{-1}$}

\newcommand{\mstar}{$M_\odot$}

\newcommand{\uchii}{{UC H{\scriptsize II}}}
\newcommand{\hii}{{H{\scriptsize II}}}

\newcommand{\Lsun}{\mbox{\,$\rm L_{\odot}$}}

\newcommand{\Tdust}{$T_\mathrm{d}$}
\newcommand{\tabnote}[1]{$^\mathrm{#1}$}

\begin{document}

   \title{Distribution and excitation of thermal methanol in 6.7 GHz maser bearing star-forming regions}

   \subtitle{I. The nearby source \object{Cepheus A}}

   \author{Karl J.E. Torstensson\inst{1,2}
          \and
          Floris F.S. van der Tak\inst{3,4}
          \and
          Huib Jan van Langevelde\inst{2,1}
          \and
          Lars E. Kristensen\inst{1}
          \and
          Wouter H.T. Vlemmings\inst{5}
          }

   \offprints{K.J.E. Torstensson}

   \institute{Leiden Observatory, Leiden University,
              PO Box 9513, 2300 RA Leiden, The Netherlands\\
              \email{[kalle;kristensen]@strw.leidenuniv.nl}
         \and
             Joint Institute for VLBI in Europe,
             PO Box 2, 7990 AA Dwingeloo, The Netherlands\\
             \email{langevelde@jive.nl}
         \and
	     	SRON Netherlands Institute for Space Research\\
	     	Landleven 12, 9747 AD Groningen, The Netherlands
	     	\email{vdtak@sron.nl}
	 	\and
	     	Kapteyn Astronomical Institute\\
	     	University of Groningen, The Netherlands
        \and
	    	Argelander-Institut f\"ur Astronomie, University of Bonn\\
	     	Auf dem H\"ugel 71, D-53121 Bonn, Germany
	     	\email{wouter@astro.uni-bonn.de}
	 }
   \date{Received September 15, 1996; accepted March 16, 1997}


\abstract
  {Candidate high mass star forming regions can be identified through
    the occurrence of 6.7 GHz methanol masers. In these sources
    	the methanol abundance of the gas must be enhanced, as the masers
	require a considerable methanol path length. The place and time of origin of this
	enhancement is not well known. Similarly, it is debated in which of the physical
	components of the high mass star forming region the masers are located.}
  {The aim of this study is to investigate the distribution and
    excitation of the methanol gas around Cep~A and to describe the physical conditions
    of the region. In addition the large
    scale abundance distribution is determined in order to
    understand the morphology and kinematics of star forming regions
    in which methanol masers occur.}
  {The spatial distribution of the methanol is studied by mapping
    line emission, as well as the column density and excitation
    temperature, which are estimated using rotation diagrams. For a
    limited number of positions the parameters are checked with non-LTE
    models. Furthermore, the distribution of the methanol abundance
    is derived in comparison with archival dust continuum maps.}
  {Methanol is detected over a  $0.3\times0.15$ pc area centred on the Cep A HW2
    source, showing an outflow signature. Most of the gas can be characterized
    by a moderately warm rotation temperature (30$-$60\,K).
    At the central position two velocity components are detected with different
    excitation characteristics, the first related to the large-scale outflow.
    The second component, uniquely detected at the central location, is probably associated 
    with the maser emission on much smaller scales of 2\arcsec.
    Detailed analysis reveals that the highest
    	densities and temperatures occur for these inner components. In the inner region 
	the dust and gas are shown to have different physical parameters.}  
  {Abundances of methanol in the range $10^{-9}$ -- $10^{-7}$ are inferred,
   with the abundance peaking at the maser position. The geometry of the 
   large-scale methanol is in accordance with previous determinations
   of the Cep~A geometry, in particular those from methanol masers. The dynamical and
   chemical time-scales are consistent with the methanol originating from a
   single driving source associated with the \object{HW2} object and the masers in its
   equatorial region.}


\titlerunning{Thermal methanol in \object{Cep A}}

\maketitle

\begin{figure*}
   \centering
   \includegraphics[angle=-90,width=17cm]{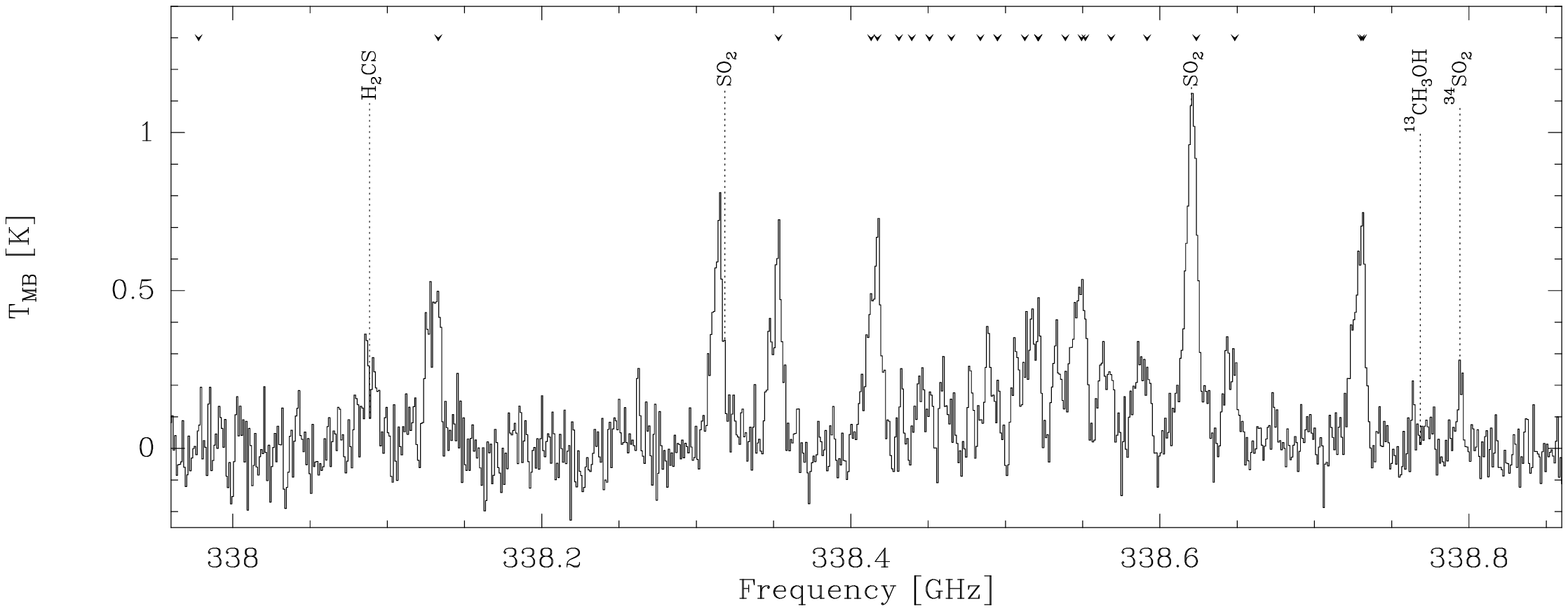}
   \includegraphics[angle=-90,width=8.3cm]{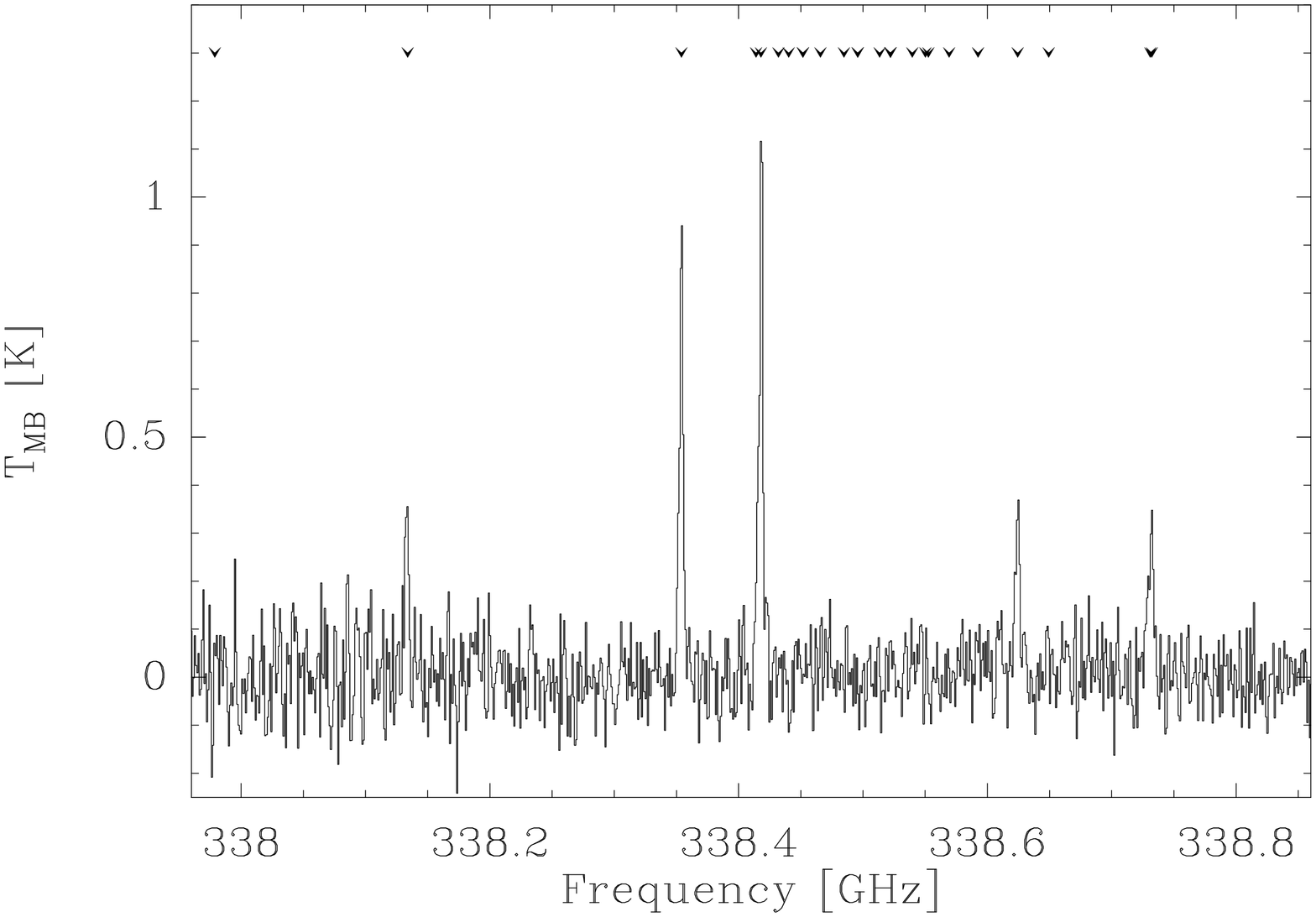}
   \includegraphics[angle=-90,width=7.7cm]{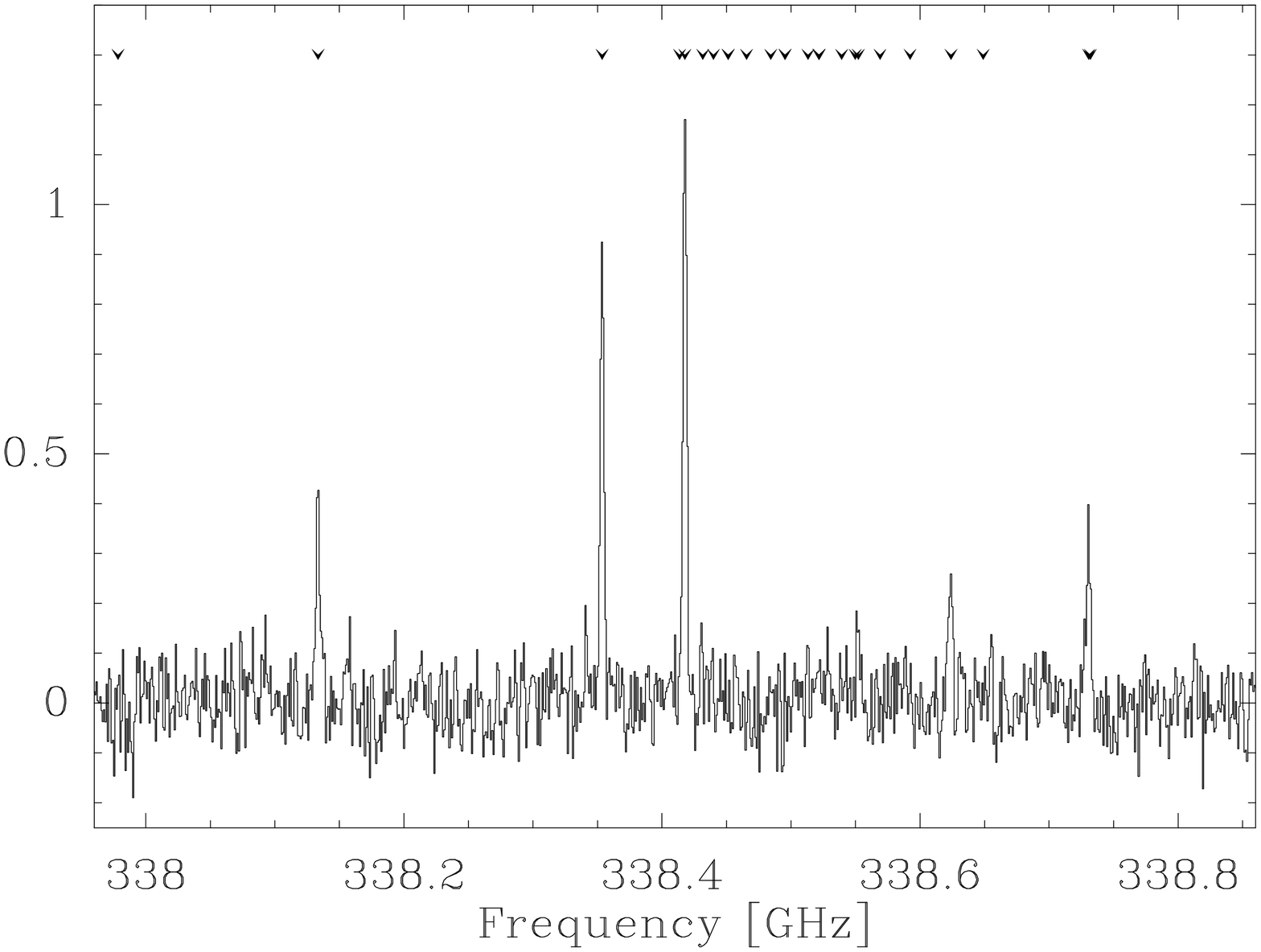}
   \caption{Sample spectra from the 3 positions identified in Sec. \ref{secdistribution}.
    {\it Top} spectrum at the "Centre" position, coincident with the \object{HW2} object.
    {\it Bottom Left} at the position of "NE Outflow" and {\it Bottom Right} "Envelope".
     The small symbols at the top indicate the frequencies of methanol transitions in this
     band.}
     \label{samplespec}
\end{figure*}

\section{Introduction}

While massive stars ($\gtrsim$10\mstar)
have an enormous impact on the evolution of the Galaxy, during their life and death, 
their birth process is still shrouded in mystery \citep[e.g.][]{kurtz05}. 
To a large extent this is due to the environment in which the high-mass stars
form. Most high mass stars are observed at large distances (compared to
low mass stars), evolve rapidly, and form deeply embedded in their
molecular clouds. Additionally, they typically form in clusters. Thus,
very high resolution data are needed to resolve them and the
disruptive influence they have on their direct environments
\citep[e.g.][]{martin-pintado05}.

Methanol maser sources are exciting targets for the detailed
study of high-mass star-formation. Continuum emission of warm dust at
sub-millimetre wavelengths has been detected at well over 95\% of the
observed 6.7~GHz methanol maser sites \citep{hill05}, while 25\% of
the same masers are associated with a detectable ultra-compact (UC)
\hii\ region \citep{walsh98}. This seems to indicate that the masers
probe a range of early phases of massive star formation and that the
methanol maser emission disappears when the \uchii\ region is created
\citep{walsh98}. 
The maser sources require high methanol abundance for their 
excitation, pointing to a recent release
of methanol into the gas phase \citep{vandertak00c}. We can thus expect to observe the 
thermal methanol lines in these sources, yielding diagnostics for the
large scale kinematics and excitation. 
Besides being signposts, masers are excellent
tracers of the geometry and small scale dynamics of these
regions. Much effort has focused on studying the kinematics and
several claims of circumstellar disks, expanding spherical shells and
jets have been presented \citep[e.g.][]{pestalozzi04, minier02,
  bartkiewicz05b, bartkiewicz09}. The evidence suggests that masers
trace an evolutionary sequence or possibly a mass range in the
formation process, or even both \citep[e.g.][]{breen10}. However, it is not clear
exactly with which of the physical structure(s) of the complex surroundings
of young massive stars they are associated. By trying to 
make a link between the maser distribution and the large scale 
methanol kinematics and
excitation, one can attempt to reach a better understanding of
the physical and chemical state of the gas of the regions in which
masers occur. 

In order to carry out such a study, we have observed
thermal methanol gas towards a sample of 15 high-mass star-forming
regions for which we have very long baseline interferometry (VLBI)
data of the 6.7~GHz methanol maser. The observations yield maps of the
distribution and physical condition of the methanol gas in relation to
the methanol maser. For one, the observations of the methanol
allow one to probe a reasonably large range of temperature and
density, due to its closely spaced $\Delta K = 0$ ($\Delta J = 1$)
transitions with a large range in excitation energies and critical
densities \citep{leurini04}. Moreover we can measure the methanol
abundance and velocity field. Clearly, the methanol abundance needs to
be significant in these regions \citep[$> 10^{-7}$,][]{sobolev97};
evaporation of \methanol\ from the dust grains in these regions, at
temperatures $\sim$100~K is thought to be responsible for this. This
specific enhancement is supposed to be so short lived that it can be
used as a chemical clock of the region \citep{vandertak00c}. The
enhancement can be used to determine the age of these regions and thus
the evolutionary stage of the star formation process.

In this paper, we describe the observations and the procedure we
have used to analyse the data. Because of its proximity and because a
wealth of data exists on this source, we will focus on
\object{Cepheus~A} East as a first example. The other sources will be
presented in a second paper.

At a distance of 700~pc \citep{moscadelli09}, \object{Cepheus A} East
is one of the closest high-mass star forming regions.  The region
has a luminosity of $2.3 \times 10^4$ \Lsun\ \citep{evans81} (scaled
to the distance of 700~pc), out of which half is thought to come from
the well-studied object \object{HW2} \citep{garay96}, which is also the
site of the 6.7~GHz methanol maser emission. The object shows both
large scale molecular outflows over $\sim$1\arcmin\ \citep{gomez99}
and signs of collapse over a similar extent
\citep{bottinelli04,sun09}. At the core of \object{HW2}, a thermal jet
extends over $\sim$2\arcsec\ and proper motion outflow velocities of
$\sim$500~\kms\ have been measured in the radio \citep{hw84, curiel06}. 
Although a superposition of gas components at different positions
cannot be completely ruled out \citep{brogan07, comito07}, the 
evidence is building that there is a disk at the centre of the
jet, observable in both molecular gas and dust \citep{patel05,
  torrelles07, jimenez-serra07}, as argued in detail in 
  \citet{jimenez-serra09}. VLBI observations of the
6.7 GHz methanol maser suggest that the methanol masers arise in a
ring-like structure, extending over $\sim$2\arcsec, straddling the
waist of \object{HW2} \citep{sugiyama08, vlemmings10, torstensson11}.

\begin{table}
\begin{minipage}[t]{\columnwidth}
\caption{\methanol\ line data for the observed transitions, adopted
from the CDMS (Cologne Database of Molecular Spectroscopy, \citet{mueller05}). 
All lines are from the $J = 7 \to
6$ band, and throughout the paper a notation like ``$-6$ E'' 
refers to the $J = 7_{-6} \to 6_{-6}$ E transition. 
Blended lines are indicated by a $^*$. }
\label{methanoldata}
\centering
\renewcommand{\footnoterule}{}  
\begin{tabular}{l l l l}
\hline \hline
Frequency   & $\mu_g^2 S_g$ & $E_u$  & Transition \\
MHz	    &      D$^2$   & K      & $k$ type \\
\hline
337969.414  	&  5.55  &  390.1   & $-$1 A $\nu_t$=1 \\ 
338124.502  	&  5.65	 &   78.1   &  0 E         \\ 
338344.628  	&  5.55	 &   70.6   & $-$1 E	   \\ 
338404.580  	&  1.49	 &  243.8   & +6 E         \\ 
338408.681  	&  5.66	 &   65.0   &  0 A         \\ 
338430.933  	&  1.50  &  253.9   & $-$6 E	   \\ 
338442.344$^*$  &  1.49	 &  258.7   & +6 A	   \\ 
338442.344$^*$  &  1.49	 &  258.7   & $-$6 A	   \\ 
338456.499  	&  2.76	 &  189.0   & $-$5 E	   \\ 
338475.290  	&  2.76	 &  201.1   & +5 E	   \\ 
338486.337$^*$  &  2.77	 &  202.9   & +5 A   	   \\ 
338486.337$^*$  &  2.77	 &  202.9   & $-$5 A	   \\ 
338504.099  	&  3.80	 &  152.9   & $-$4 E	   \\ 
338512.627$^*$  &  3.81	 &  145.3   & $-$4 A	   \\ 
338512.639$^*$  &  3.81	 &  145.3   & +4 A	   \\ 
338512.856$^*$  &  5.23	 &  102.7   & $-$2 A	   \\ 
338530.249  	&  3.82	 &  161.0   & +4 E	   \\ 
338540.795$^*$  &  4.60	 &  114.8   & +3 A	   \\ 
338543.204$^*$  &  4.60	 &  114.8   & $-$3 A	   \\ 
338559.928  	&  4.64	 &  127.7   & $-$3 E	   \\ 
338583.195  	&  4.62	 &  112.7   & +3 E	   \\ 
338614.999  	&  5.68	 &   86.1   & +1 E	   \\ 
338639.939  	&  5.23	 &  102.7   & +2 A	   \\ 
338721.630  	&  5.14	 &   87.3   & +2 E	   \\ 
338722.940  	&  5.20	 &   90.9   & $-$2 E	   \\ 
\hline	       
\end{tabular}
\end{minipage}
\end{table}

In Sec.~\ref{secobs} we describe the observations and 
calibration methods. The results are presented in Sec.~\ref{secresults}
in the form of a number of sample spectra and maps of integrated
line strength and velocity. In order to study the large scale methanol
excitation we construct rotation diagrams and produce maps of the
derived physical quantities in Sec.~\ref{secanalysis}. We present 
non-LTE calculations for a number of positions to evaluate the 
limitations of the diagram analysis in Sec.~\ref{secnonlte}.
Subsequently we discuss estimates of the hydrogen column density
before we present the methanol abundance in Sec.~\ref{secabundance}.
The relation of our findings to the \object{Cep A} star forming process and
in particular the occurrence of the methanol maser are discussed
in Sec.~\ref{secdiscussion}.

\section{Observations and data reduction}\label{secobs}

The JCMT\footnote{The James Clerk Maxwell Telescope is operated by the
  Joint Astronomy Centre on behalf of the Science and Technology
  Facilities Council of the United Kingdom, the Netherlands
  Organisation for Scientific Research, and the National Research
  Council of Canada.} observations of \object{Cep A} East
(\object{HW2}) (22$^h$56$^m$17.9$^s$ +62\degr01\arcmin49\arcsec\
(J2000)) were performed on June 2 and June 17, 2007. We used the array
receiver HARP, which has 16 receptors with a spacing of
$\sim$30\arcsec\ in a 4$\times$4 grid, resulting in a footprint of
$\sim$2\arcmin. The receptors are single side-band receivers with a
sideband rejection of $>$10~dB. We used the observing mode HARP5, a
type of beam-switching (jiggle-chop) mode where several short scans at
different positions of the target source (10~s per jiggle position)
are observed before switching to the off position. As a result,
several on-source observations share the same off-source observation
which minimises the overhead while still providing good baselines. The
observing mode results in a map of $20 \times 20$ pixels covering
2\arcmin$\times$2\arcmin, thus each pixel is 6\arcsec x6\arcsec. This
ensures proper Nyquist sampling of the 14\arcsec\ telescope beam at
338~GHz. 

The ACSIS correlator backend was set up with a 1~GHz (880~\kms)
bandwidth and 2048 channels centred on the methanol $7_{0} \to
6_{0}$ A$^+$ line at 338.41~GHz. The frequency set up allows us to
cover 25 \methanol\ lines (Table \ref{methanoldata}) with a velocity
resolution of $\sim$0.43~\kms\ (488~kHz). In total we have an
effective (on source) integration time of $\sim$5 min per pixel with a
typical system temperature ($T_\mathrm{sys}$) of $\sim$280~K. For our
analysis we have adopted a main beam efficiency
($\mathrm{\eta}_{\mathrm{mb}}$) of 0.6. Regular pointings were done of
the normal calibrators and we estimate to have an absolute pointing
accuracy of $\approx$1\arcsec.

We used the Starlink package (Gaia and Splat) for initial inspection
of the data. It was found that receptor R06 suffered from bad
baselines (large ripples is some scans). For that receptor only these
poor scans were removed, after which the scans were combined. The data
were then converted to GILDAS/CLASS format and the remaining data
reduction and analysis was performed in CLASS. A linear baseline was
fitted to several spectral regions without any line emission and
subtracted from the spectra. Next, to improve the signal to noise, the
data were smoothed in the spectral domain resulting in a velocity
resolution of 0.87~\kms\ and an rms of $\sim65$~mK.

The analysis was performed on a pixel by pixel basis, by first fitting
a Gaussian to the strongest unblended line ($7_{-1} \to 6_{-1}$ E) in
the spectra. The measured velocity and width of this line was then
used to place windows around all other features in the spectra, for which
the emission was integrated. Due to some overlap between different
lines, only lines with a $>5\sigma$ ($\sim0.6$~K\,\kms) detection are
included in the subsequent analysis. In the centre area, at the
position of HW2, a second velocity 
component can be distinguished. 
Where the two velocity components
could be separated, manual analysis by fitting individual Gaussians to
the lines was performed.

\section{Results}\label{secresults}

\subsection{Methanol lines}

In the following we refer to each pixel in the map with its
respective offset coordinates ($\Delta \alpha$, $\Delta \delta$),
measured in arc seconds, with respect to the centre of the map (J2000
22$^h$56$^m$17.88$^s$ +62\degr01\arcmin49.2\arcsec).
Fig.~\ref{samplespec} shows the observed spectrum at 3 positions:
(+3\arcsec,3\arcsec) ``Centre'', (+27\arcsec,+9\arcsec) ``NE outflow'',
(+21\arcsec,-3\arcsec) ``Envelope'', where the designations can be
understood from the forthcoming discussion in
Sec. \ref{secdistribution}. Focusing at the centre spectrum, all but
four of the detected lines are due to methanol (indicated by ticks at
the top); the others are identified as SO$_2$, $^{34}$SO$_2$ and H$_2$CS.
The \methanol\ K=+1 E line
is a blend with another SO$_2$ line, also indicated in the figure. 

\begin{figure}
   \centering
   \includegraphics[width=8cm]{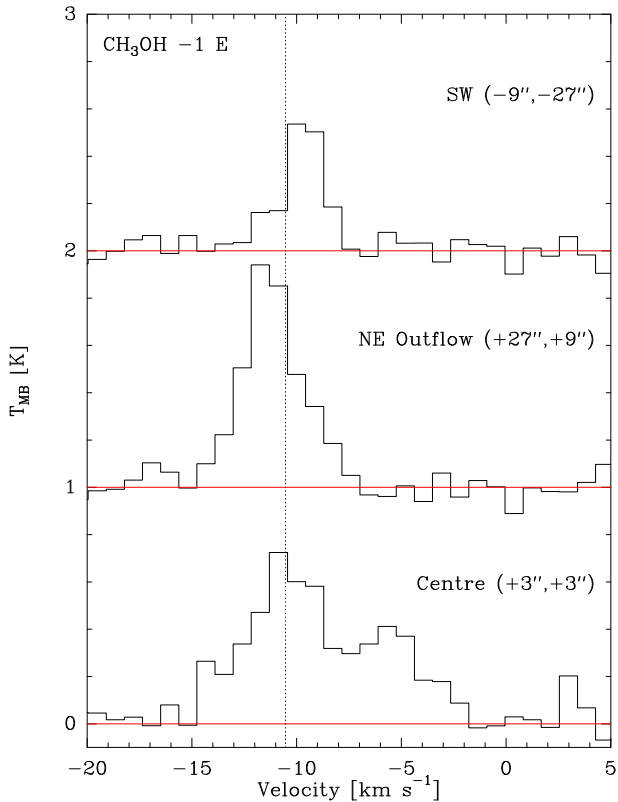}
   \caption{Velocity profiles of the -1 E line at 3 positions. Although the profiles are 
   Gaussian in the outflow, there is a clear shift in the velocity. At the ``Centre'' position
   there are clearly two velocity components present. The dotted line is taken to be the
   system velocity, corresponding to the velocity of the dominant component at the ``Centre'' 
   position.}
         \label{specblowup}
\end{figure}

In the centre spectrum, two gas components are observed: the main 
component with $v_{\rm lsr} = -10.5$\,\kms\ and a secondary red shifted at $-5.2$\,\kms,
as can be seen in the profile of the $-$1~E transition (Fig. \ref{specblowup}).
We note that the red-shifted velocity profile is overlapping with the range spanned
by the methanol masers, which are observed with velocities between
$v_{\rm lsr} = -4.6$ and $-1.5${\kms} \citep{torstensson11}.
The red-shifted component is most readily identified in the higher-K
transitions. While the first component is brightest in the low-K
transitions and not detected in the higher-K lines, the $-5.2$\,\kms\ component
is brighter in the higher-K lines and less bright in the
lower-K lines. For the K=3 transitions, the two components
are of similar strength. There
is a weak detection of the methanol isotopologue $^{13}$\methanol\ $J = 13 \to 12, \Delta k=+1$
in this component, indicating optical depth effects may be important for this.
Also careful analysis, including averaging
over a few pixels, shows that the torsionally excited $-1$ A v$_\mathrm{t}$=1 
line at 337.969 GHz is detected in the ``Centre'' region in the $-5.2$\,\kms\ velocity component.

Otherwise the observed line profile appears Gaussian in all components and
does not show evidence of blue/redshifted wings. However, a modest 
velocity shift of the line is observed
across the source with lower velocities to the NE and higher
velocities to the SW (Fig. \ref{specblowup}).

In order to start a consistent analysis of all methanol line emission
in our data cube, we have extracted the integrated line flux from each
line as described above (Sec.~\ref{secobs}). To illustrate our
procedure, we list the integrated line emission of the spectral
features for the main velocity component at the ``Centre'' position 
from Fig.~\ref{samplespec} in Table~\ref{samplefits}. For lines
that are blended we have chosen a simple strategy. If the
strengths ($S_g\mu_g^2$) of the lines are similar and they have comparable
upper energy levels, we have split the measured flux between the two
lines. If however one line has a significantly higher line strength
and lower upper energy level we attribute all the emission to the
strong line.
In the case of the blend between the K=$-$2 A and K=$\pm$4 A,
where the lines have comparable line strengths and energy levels we
assume the total measured flux as an upper limit to all the lines.

\begin{table}
\begin{minipage}[t]{\columnwidth}
\caption{Line parameters for the main velocity component measured at the position of \object{HW2} (+3\arcsec,+3\arcsec)} 
\label{samplefits}      
\centering                          
\renewcommand{\footnoterule}{}  
\begin{tabular}{l l l }        
\hline\hline                 
Line		& Flux	& Width \\
K, type\footnote{See Table 1 for transition details} &  K\,\kms  & \kms\\
\hline
0 E		& 2.16	& 4.21	\\ 
$-$1 E	& 3.30	& 4.79	\\ 
0 A		& 3.60	& 4.79	\\ 
$-$6 E	& 0.03	& 1.74	\\ 
+6 A	& 0.29	& 3.80	\\ 
$-$6 A	& 0.29	& 3.80	\\ 
$-$5 E	& 0.61	& 4.81	\\ 
+5 E	& 0.31	& 3.15	\\ 
+5 A	& 0.36	& 4.18	\\ 
$-$5 A \footnote{Blend with +5 A}	& 0.36	& 4.18	\\ 
$-$4 E	& 1.05	& 4.79	\\ 
$-$2 A\footnote{Blend with $\pm$4 A}	& 2.00	& 4.79	\\ 
+4 E	& 1.12	& 4.79	\\ 
+3 A	& 1.37	& 4.79	\\ 
$-$3 A\footnote{Blend with +3 A}	& 1.37	& 4.79	\\ 
$-$3 E	& 0.93	& 4.07	\\ 
+3 E	& 0.94	& 4.97	\\ 
+2 A	& 1.42	& 5.38	\\ 
+2 E	& 1.55	& 4.38	\\ 
$-$2 E\footnote{Blend with +2 E}	& 1.55	& 4.38	\\ 
\hline                                   
\end{tabular}
\end{minipage}
\end{table}

\subsection{Spatial distribution of methanol}\label{secdistribution}

\begin{figure}
   \includegraphics[width=8cm]{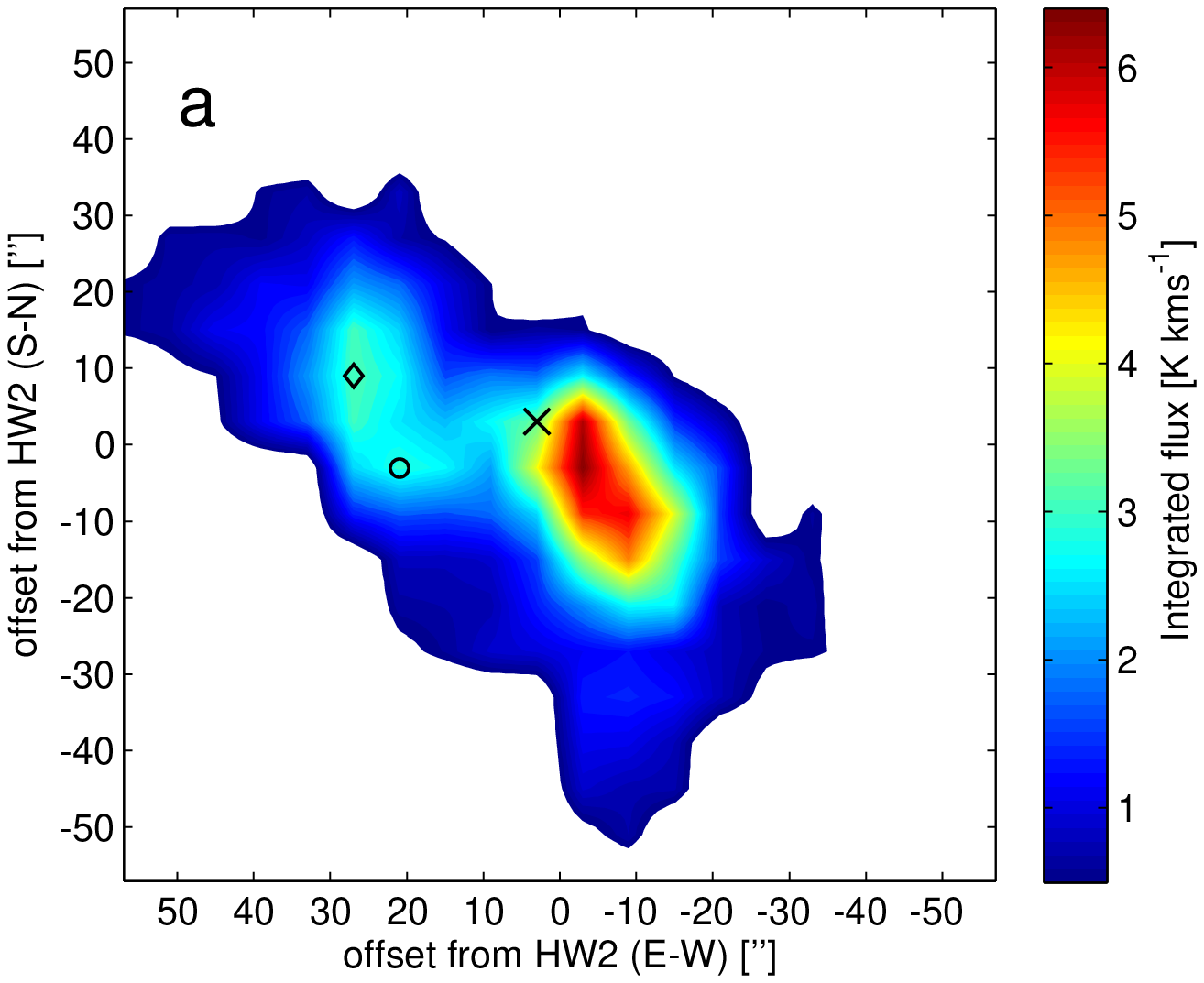}
   \includegraphics[width=8cm]{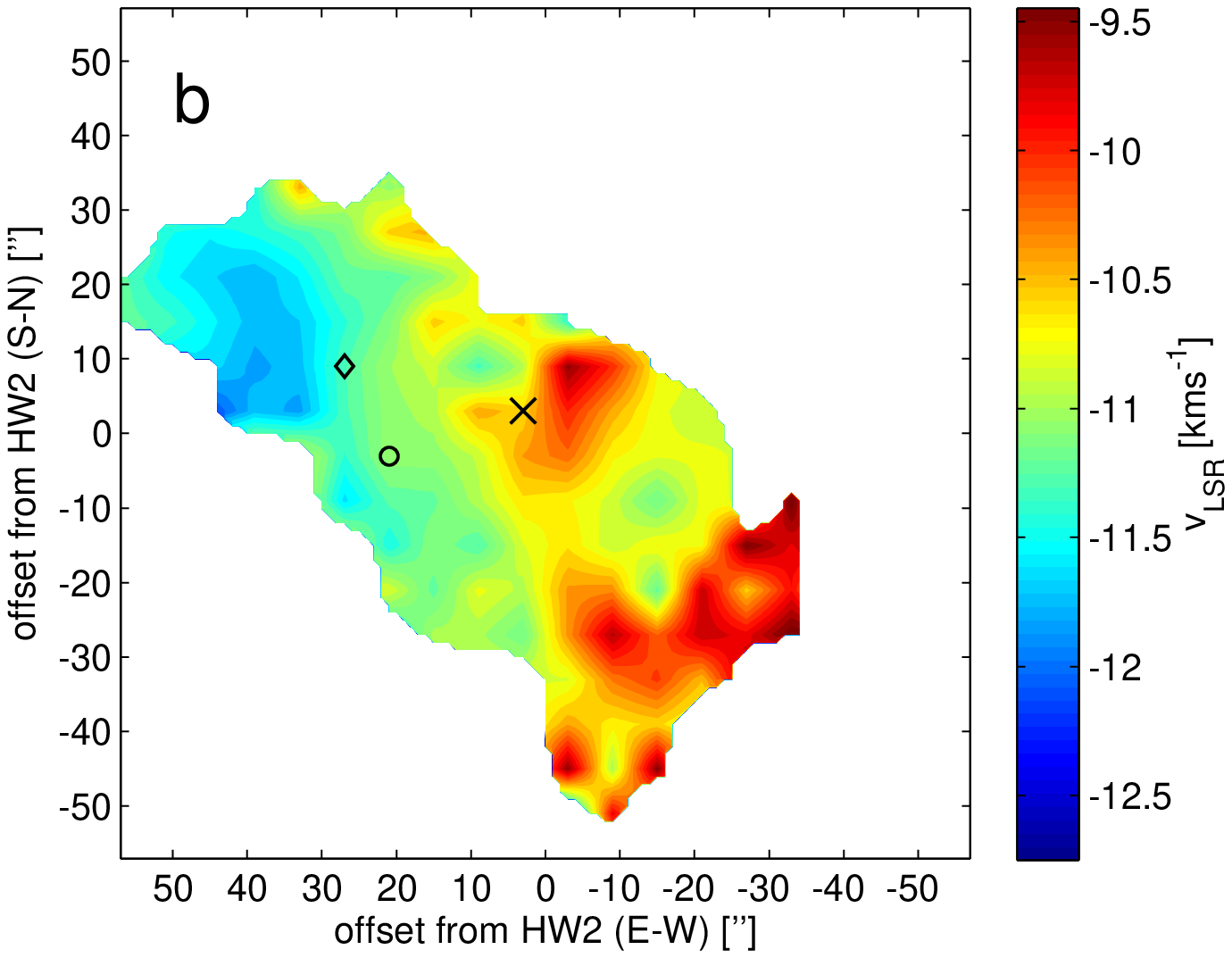}
   \includegraphics[width=8cm]{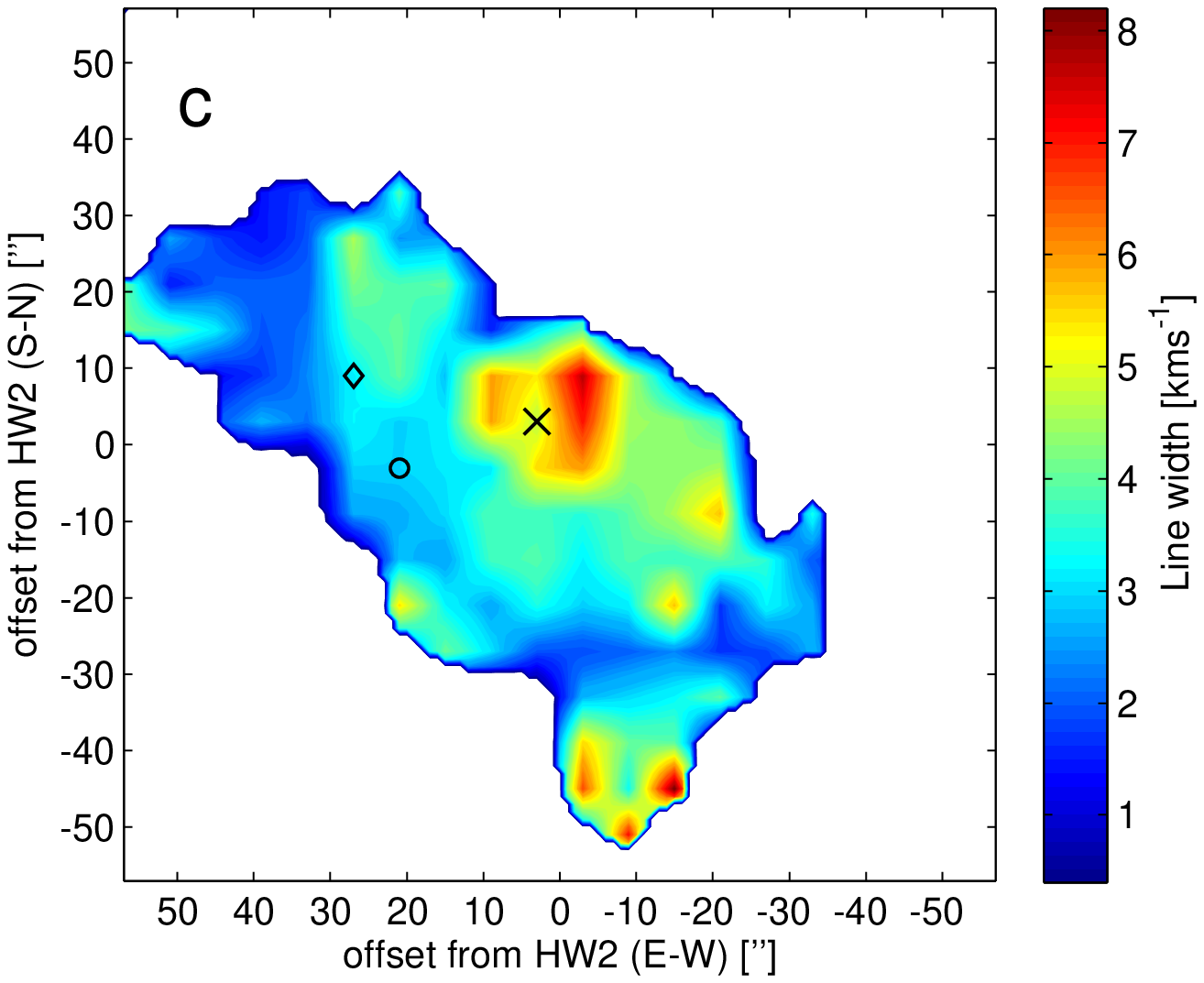}
   \caption{From top to bottom (a-c) the integrated flux, the
      velocity field, and the line width, of the methanol $7_{-1}
      \to 6_{-1}$ E-type transition, the strongest unblended
      line in the observed spectra. In all maps the cross marks the ``Centre'' position,
      the diamond ``NE Outflow'' and the circle indicates the position
      where the ``Envelope'' spectra are evaluated. }
    \label{line03plots}
\end{figure}

Fig.~\ref{line03plots} shows the spatial distribution of the flux,
the central velocity, and the width of the methanol $7_{-1} \to
6_{-1}$ E line, which is the brightest unblended feature in the
spectra. The methanol maser emission arises just to the NE of the
centre of the map (+0.7\arcsec, +0.4\arcsec) and is indicated with a
cross in all the figures. The white area in the maps represents
regions where the line flux is less than our $5\sigma$ detection limit
of 0.6~K\,\kms.

The line emission is constrained to an area, measured from the points
furthest separated in the intensity map (Fig.~\ref{line03plots}a), with
an extent of $\sim$85\arcsec\ by 43\arcsec\ ($\sim$0.29~pc by
$\sim$0.15~pc), and is oriented in the NE-SW direction. The velocity
map (Fig.~\ref{line03plots}b) shows that the methanol gas is entrained
in an outflow with the blue-shifted emission to the NE and the
red-shifted emission to the SW. Measured between the highest and
lowest velocities in the map, the outflow has a modest velocity
gradient of $\sim$3.1~\kms\ over $\sim$0.20~pc. The integrated line
flux peaks in the centre of the map at $\sim5.2$~K\,\kms, with a
bright extension to the SW over $\sim$13\arcsec ($\sim$0.044~pc),
which is associated with the red-shifted emission. Both the velocity
field and line width maps (Figs. \ref{line03plots}b and c) show a
maximum in the centre of the map to the N of HW2. This is due to
blending of the two velocity components that we are unable to
decompose. A second maximum in the integrated line flux is observed
$\sim$28\arcsec ($\sim$0.095~pc) to the NE of the centre, which is
associated with the blue shifted part of the emission and a secondary
peak in the line width map. Throughout the paper we will refer to this
position as the ``NE outflow''. The quiescent methanol emission that
is not associated with the outflows and does not show any peak in the
integrated line flux map we will refer to as the ``Envelope''.

\section{Analysis}\label{secanalysis}

\subsection{Rotation diagrams}\label{rotsection}

From the above spectral cuts in 3 positions and the brightness distribution of
one of the lines, it is clear that the observations trace a complex
region in which the dynamic and excitation characteristics change, at
least on the scales represented by the beam size of the telescope.
To make a representation of the large scale distribution and
excitation of methanol, we have created rotation diagrams (Boltzmann
plots) at every pixel in order to be able to study the characteristics of the
excitation distribution. Inherent to the method are a number of
assumptions, namely that the gas at each velocity and spatial resolution element
can be described by a single
excitation temperature ($T_\mathrm{rot} \equiv T_\mathrm{ex}$), that
all the lines are optically thin, i.e. $\tau<<1$, and that the size of
the emitting region is the same for all lines. Although not all these
assumptions may hold for all positions, the strength of the method
is that it shows general trends on the relevant scales of our
observations. In Sec. \ref{secnonlte} we will test the validity of
 these assumptions.

For each pixel at which we measure at least three methanol transitions
with S/N$>$5, we plot the upper energy level $E_u$ [K] of the
transitions on the x-axis versus the logarithm of the column density
in the upper energy state divided by the statistical weight of that
upper level $\log(N_\mathrm{u}/g_\mathrm{u})$ on the y-axis. The
weighted column density ($N_u/g_u$) can be calculated with Eq. \ref{eqrotdiag}
\citep{helmich94} and the appropriate coefficients in Table
\ref{methanoldata}. In the equation $Q(T_\mathrm{rot})$ is the
partition function, $\mu$ the permanent dipole moment [Debye], $S_g$
the line strength value from \citep{blake87}, and $T_\mathrm{MB}$ the
main beam temperature is the antenna temperature scaled by the main
beam efficiency.

\begin{eqnarray}
\frac{N_u}{g_u} &=& \frac{N_\mathrm{M}}{Q(T_\mathrm{rot})} e^{-E_u/T_{rot}} \approx \frac{1.67\cdot10^{14}}{\nu \mu_{g}^2 
S_{g}} \int T_\mathrm{MB}dv
\end{eqnarray}\label{eqrotdiag}

By fitting a straight line through these data points we determine the
rotation temperature $T_\mathrm{rot}$ [K] and column density
$N_\mathrm{M}$ [cm$^{-2}$] of the methanol gas. The rotation
temperature is the negative inverse of the slope of the fitted line
and the total column density is where the extrapolated line crosses
the y-axis.

\begin{figure*}
   \centering
   \includegraphics[width=8cm]{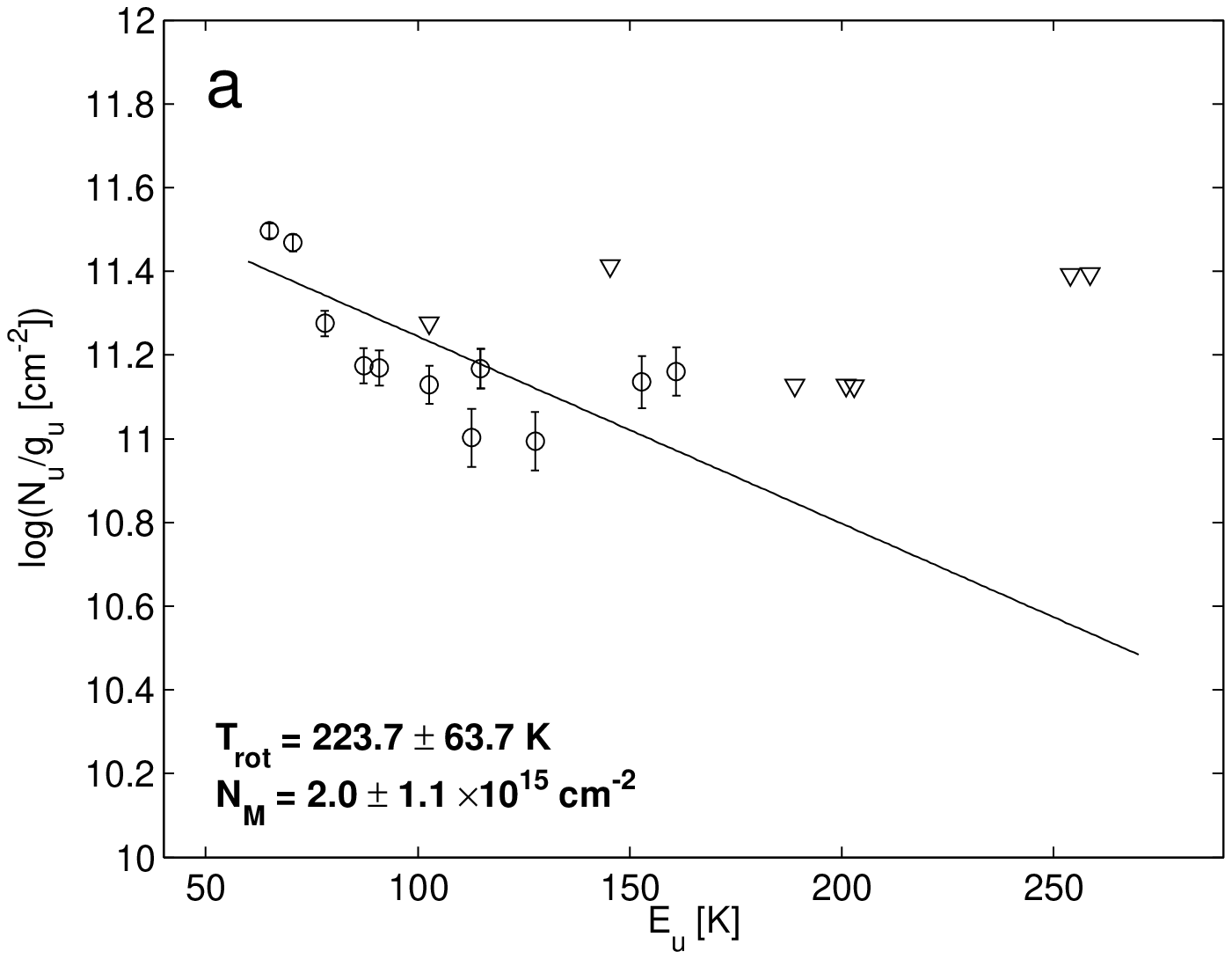} 
   \includegraphics[width=8cm]{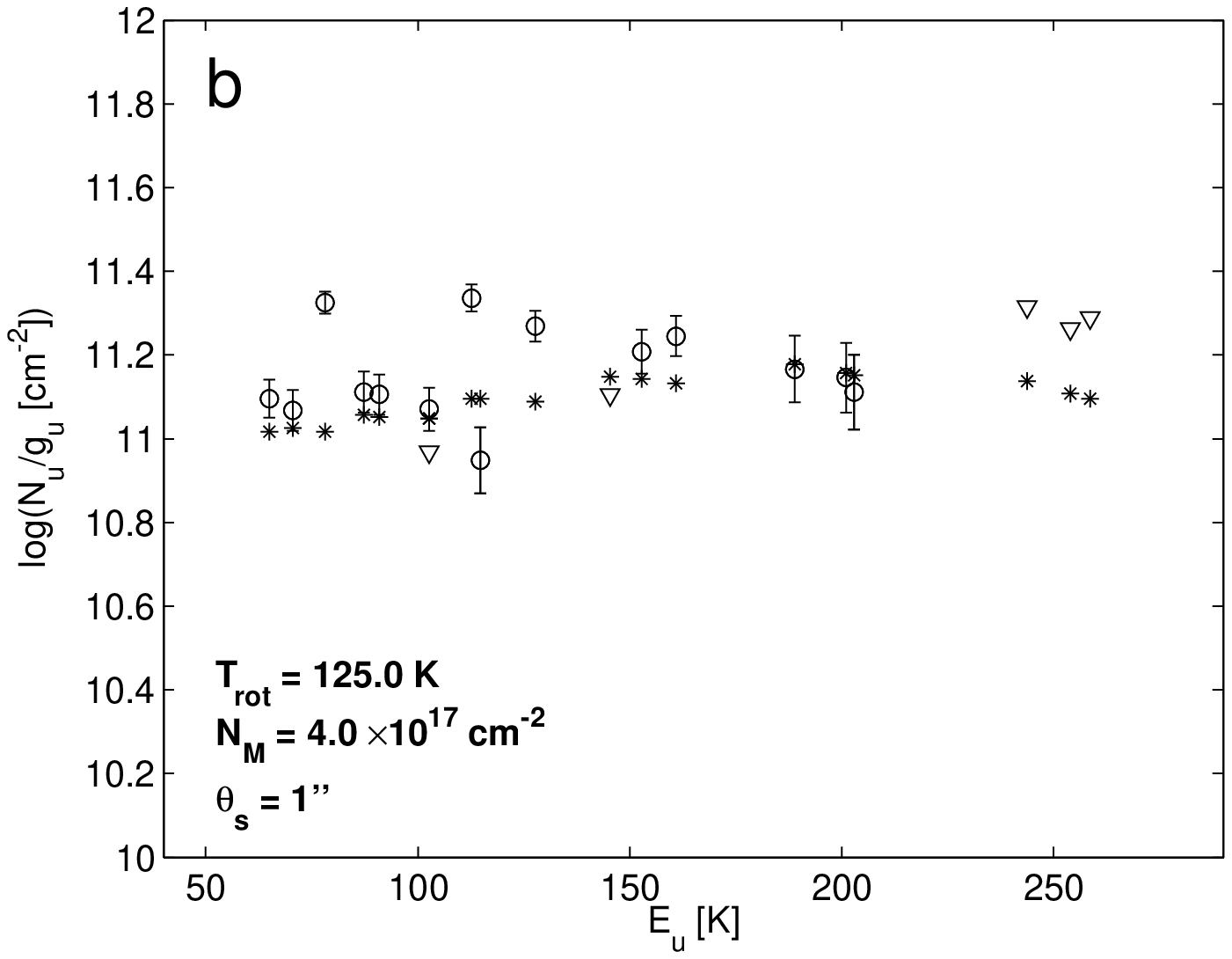} 
   \includegraphics[width=8cm]{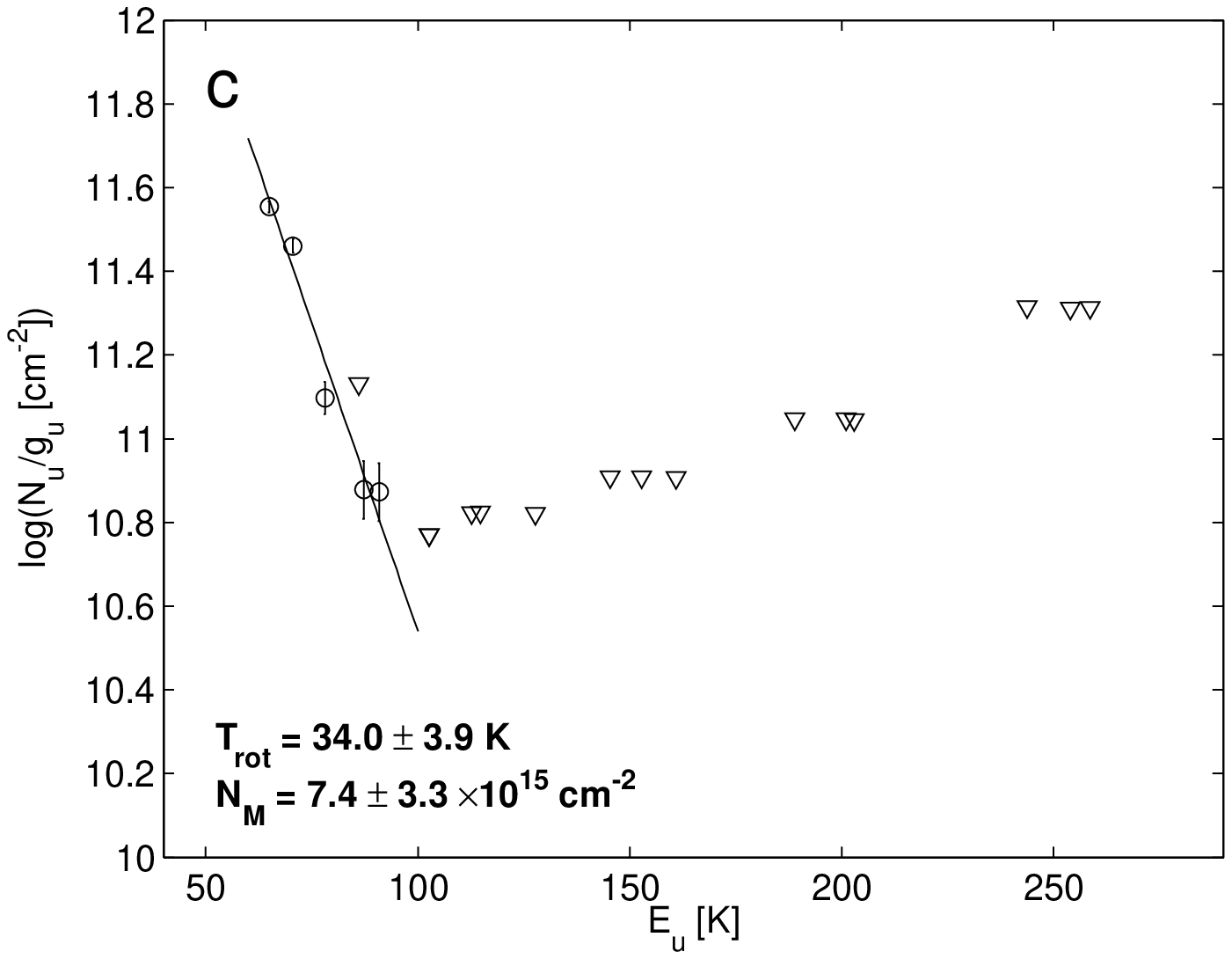} 
   \includegraphics[width=8cm]{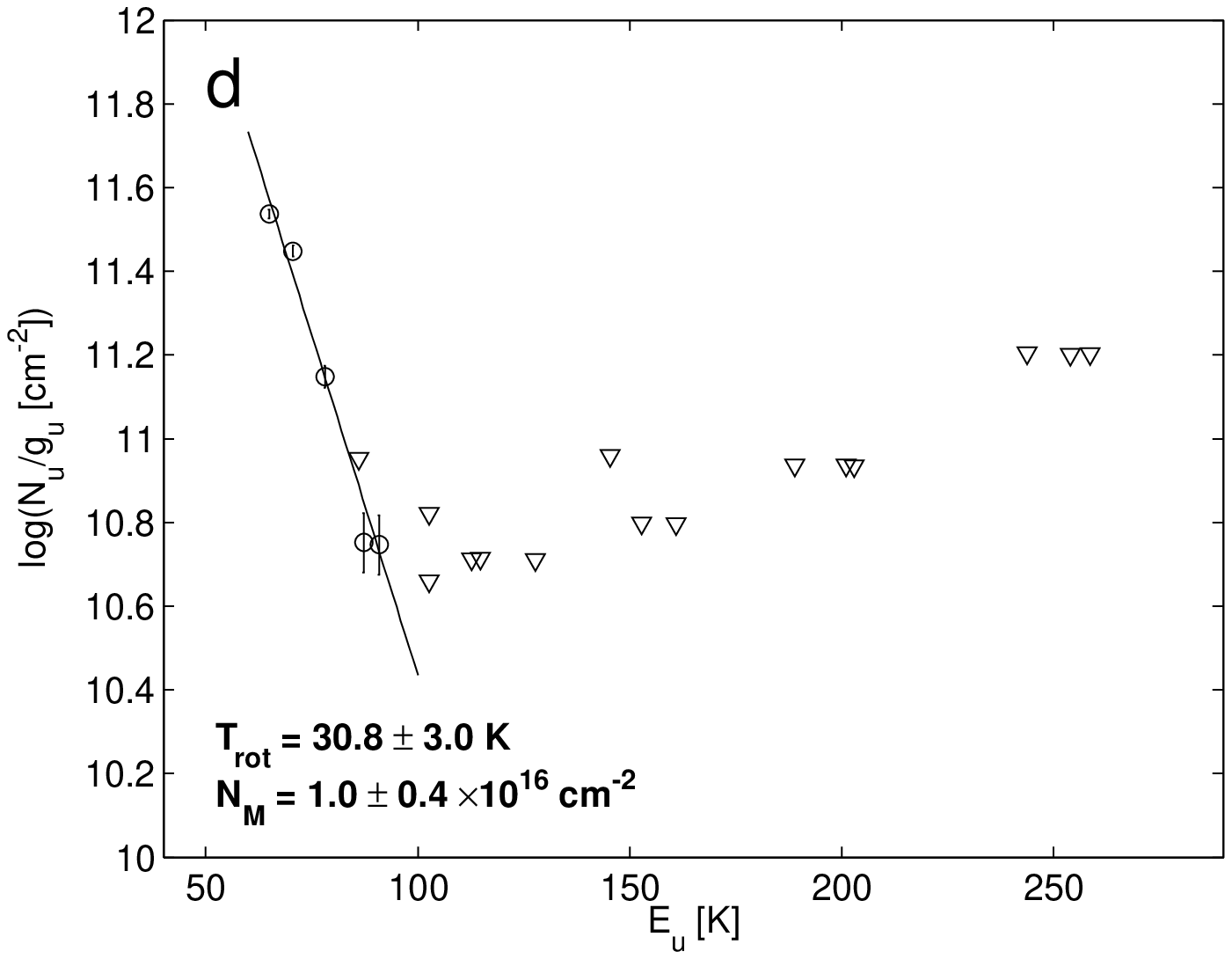} %
      \caption{Rotation diagrams for three specific components and 
      a population diagram for the fourth component identified in Sec. \ref{secdistribution}. 
      Triangles indicate upper limits and error bars the 1$\sigma$ standard deviation 
      in the flux determination. The diagrams displayed are: 
      {\it a}: rotation diagram for the main, $-10.5$\,\kms\ velocity component at the "Centre" position (+3\arcsec,+3\arcsec). 
      {\it b}: population diagram for the red-shifted, $-5.2$\,\kms\ velocity component at the "Centre" position (+3\arcsec,+3\arcsec);
      for this model asterisks indicate a best fit that includes optical depth effects and beam dilution 
      (see Sec.~\ref{secpopmaps}),
      {\it c}: rotation diagram for the ``NE Outflow'' (+27\arcsec,+9\arcsec), 
      {\it d}: rotation diagram for ``Envelope'' (+21\arcsec,-3\arcsec). }
      \label{rotdiacomb}
\end{figure*}

Fig.~\ref{rotdiacomb} shows the resulting rotation diagrams of the
``Envelope'' (the lower temperature quiescent gas), the ``NE
outflow'', and of the first velocity component at the ``Centre''
position of HW2. For the two positions in the outflow and envelope a
single gas component provides a good fit to the data points with a
rotation temperature of 30~K to 35~K and a column density of
$10^{16}$~cm$^{-2}$. However, in the centre region things are more
complex. The velocity component at $-10.5$\,\kms\ can be fitted by a straight
line, yielding a rotation temperature of 224~K and a column density of
2$\times 10^{15}$cm$^{-2}$. For the $-5.2$\,\kms\ velocity component,
the points lie nearly horizontal in the diagram and we would only able to put 
a lower limit of $>$300\,K on the rotation temperature with the simple fit.

\subsection{Population diagram modeling}\label{secpopmaps}

Clearly the basic assumptions going into the rotation diagram analysis 
are violated for the ``Centre'' position and a
successful explanation must accommodate both an apparent negative rotation
temperature and the relative weakness of the low K lines for the red-shifted component. 
Non-LTE conditions, optical depth and beam-dilution effects can all play a 
role here.
As a first step, to investigate the effect of line optical depth on the above results, we have
adopted the method of \citet{goldsmith99} which includes the source size ($\theta_s$) as 
a free parameter and allows for optical depth effects. We have performed a $\chi^2$ analysis for 
$10<T_{\rm rot}<500$~K, $10^{14}< N_M <10^{18}$~cm$^{-2}$, and  
$1\arcsec\ \le \theta_s \le 14\arcsec$, i.e. a beam dilution factor between 200 and 1.
For the red-shifted component, we find the 
best fit to be for a rotation temperature of 125~K, a column density of 
$4\times 10^{17}$~cm$^{-2}$, and $\theta_s = 1\arcsec$, much smaller 
than the beam. The merit of this model can be seen in panel Fig.~\ref{rotdiacomb}b. 
In this specific case the maximum optical depth encountered
was $\tau=6.8$. As this seems to be the component with the highest column density it is not 
surprising that $^{13}$\methanol\ is detected in this component.
Obviously, the highly excited K lines lead the model towards a high column density, 
but at the same time, the only way for the model to accommodate the lower K lines by 
increasing the optical depth is to decrease the source size. It should however be 
noted that this analysis assumes a single excitation temperature. We will discuss 
non-LTE modeling in Sec.~\ref{secnonlte}.

\subsection{Spatial distribution of the excitation}\label{secrotmaps}

\begin{figure}
   \centering
   \includegraphics[width=8cm]{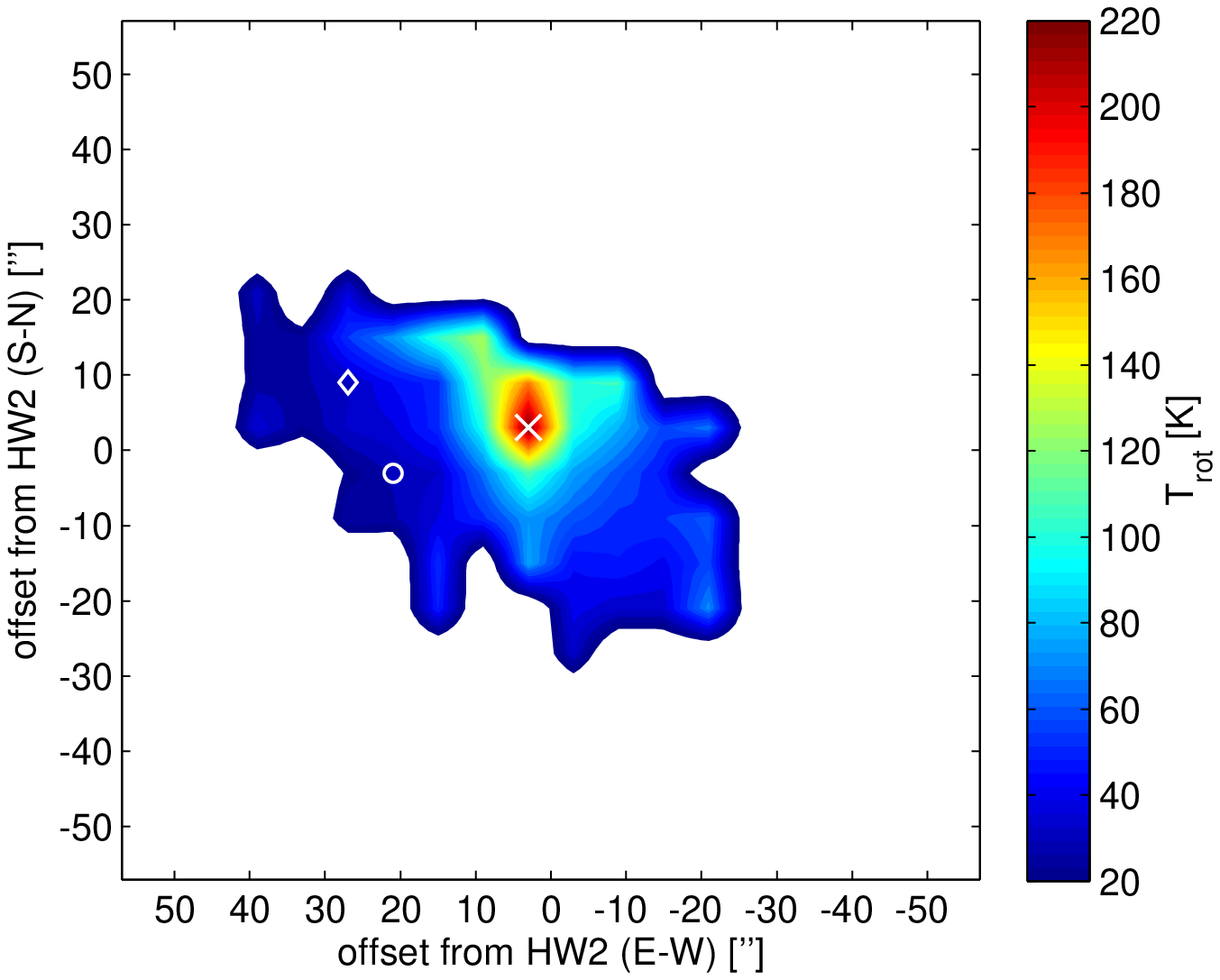}
      \caption{Map of derived rotation temperatures. Symbols are as in Fig~\ref{line03plots}.}
         \label{Trotmap}

   \centering
   \includegraphics[width=8cm]{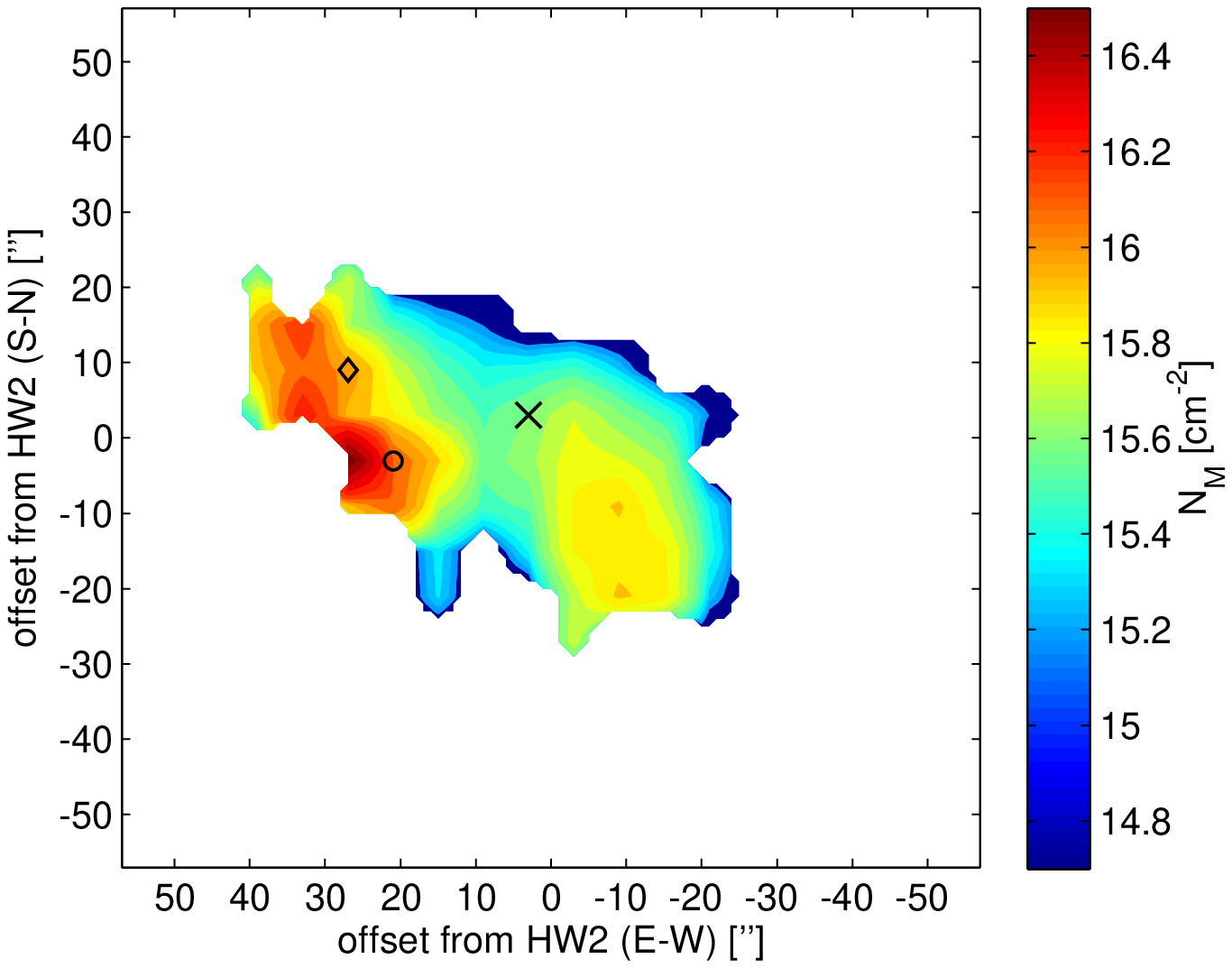}
      \caption{Map of derived column densities. Symbols are as in Fig~\ref{line03plots}.}
         \label{Nrotmap}
\end{figure}

Maps of the derived rotation temperatures and column densities are
presented in Figs.~\ref{Trotmap} and \ref{Nrotmap}. In the maps
we have excluded the results of the $-5.2$\,\kms velocity component seen at
the position of \object{HW2}. 
The methanol rotation temperature ranges from $\sim$20~K
to $\sim$100~K over most of the map, but shows a pronounced peak of
$\gtrsim$200~K close to the \object{HW2} position. We find column densities
between $10^{15}$~cm$^{-2}$ and $3 \times 10^{16}$~cm$^{-2}$. Overall, 
the methanol column density distribution is quite smooth, showing a minimum towards the centre position. 
In Sec. \ref{secnonlte} we will investigate whether this minimum is an 
artifact of the rotation diagram method. 
Additionally, some care must be taken
when analysing the results in the centre region. Only at the position
of \object{HW2} are we able to separate the two velocity components of
the gas. In the neighbouring pixels the integrated line fluxes are
therefore overestimated due to blending.

\subsection{Non-LTE analysis}\label{secnonlte}

\subsubsection{Method}

In order to check the validity of the assumptions of the rotation diagram
analysis, we have run a set of non-LTE models. By comparing the resulting
synthetic spectra with the observed spectra, these models also allow us
to investigate the physical conditions (i.e. temperature and density)
at distinct positions in the methanol emitting region. We have used the
RADEX\footnote{http://www.strw.leidenuniv.nl/$\sim$moldata/radex.html}
package, that uses statistical equilibrium for the population
calculations and an escape probability method to calculate the optical
depth \citep{vandertak07}. The main assumptions are that the medium is
uniform and that the optical depth is not too high ($\lesssim
100$). The most accurate collisional rates are available for the first 100 levels
of methanol \citep{pottage04,schoeier05} and we therefore restrict this
part of the analysis to the 24 relevant lines from Table~\ref{methanoldata}.

Initial runs of the non-LTE model indicated that the methanol
excitation is more strongly governed by the density than the 
temperature. Using a uniform sphere geometry, we considered
kinetic temperatures between 30 and 300~K and \hmol\
densities between $10^4$ and $10^7$~cm$^{-3}$. Even at the 
lowest temperature all the lines, except
the K=$\pm$5 and K=$\pm$6, are excited in the high-density case. On
the other hand, at low-density and high-temperature, only the lowest
three K levels (K=0,$\pm$1,$+$2) are populated. A combination
of high temperature and high density is required to excite all lines,
making this methanol band useful as a density and temperature tracer
\citep{leurini04}. 

Starting with model values comparable to the results from Sec. \ref{rotsection},
it is possible to model the observed line strengths of the large-scale
emission with the simple assumption that the emitting region is filling the beam.
For these parameters all the lines are optically thin ($\tau$ < 0.3), validating the
assumption made in the rotation diagram analysis for the
large scale emission.

In the above calculations the background radiation field was represented
by the cosmic microwave background with $T_{\mathrm{bg}}$=2.73~K. Models with background radiation temperatures in the range 3 -- 300 K were run,
but this has no clear impact on the lines considered
here. Mostly the contrast of the lines with respect to the background 
was reduced without changing their relative intensities. This result can 
be understood from the fact that our models consider only the torsional-vibrational ground state of methanol 
\citep{leurini07}. These models are inconclusive about
the influence of the radiation field, which are expected to be
important in the maser region.

\subsubsection{Non-LTE results}\label{secnonlteresults}

\begin{figure*}
  \centering
  \includegraphics[width=8cm]{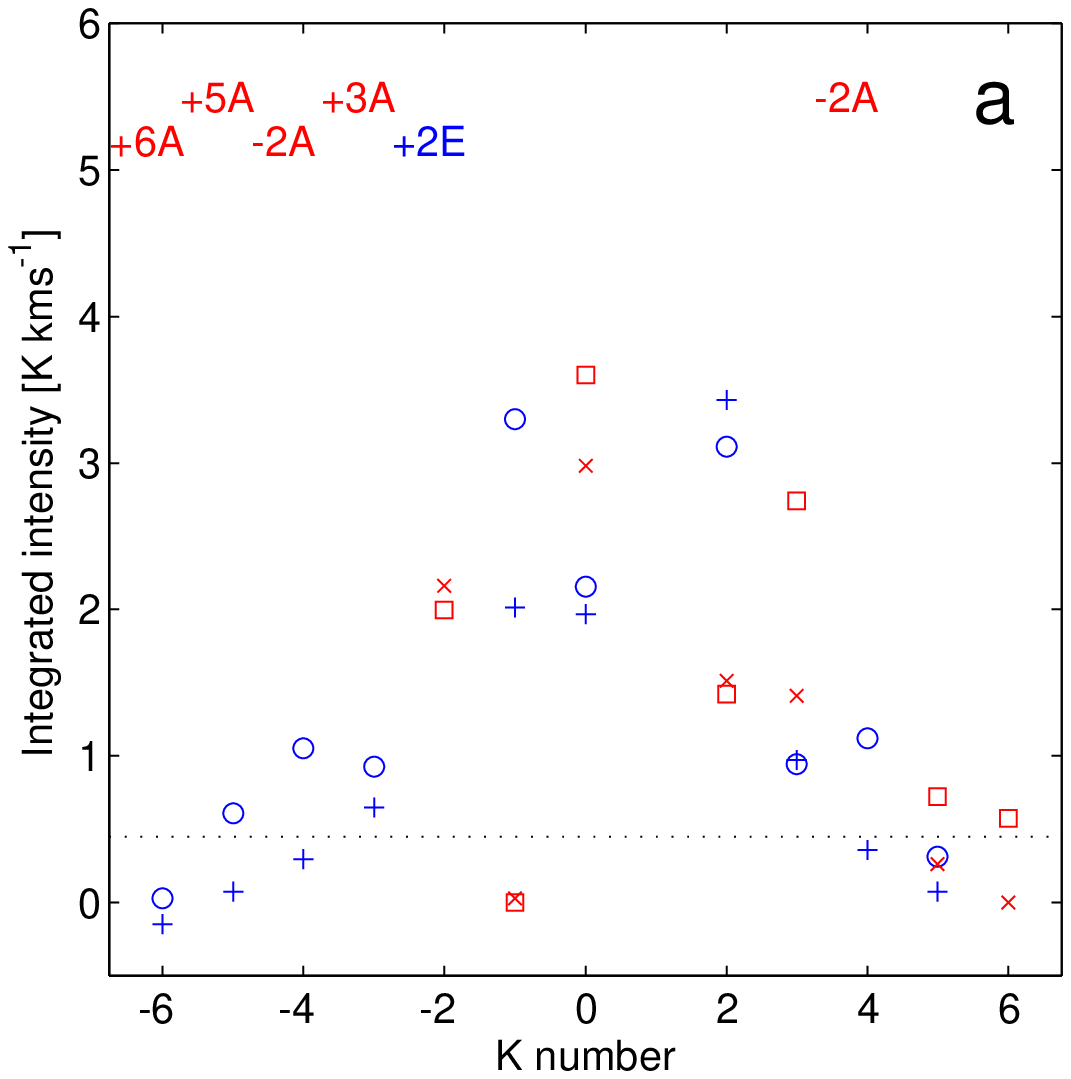}
  \includegraphics[width=8cm]{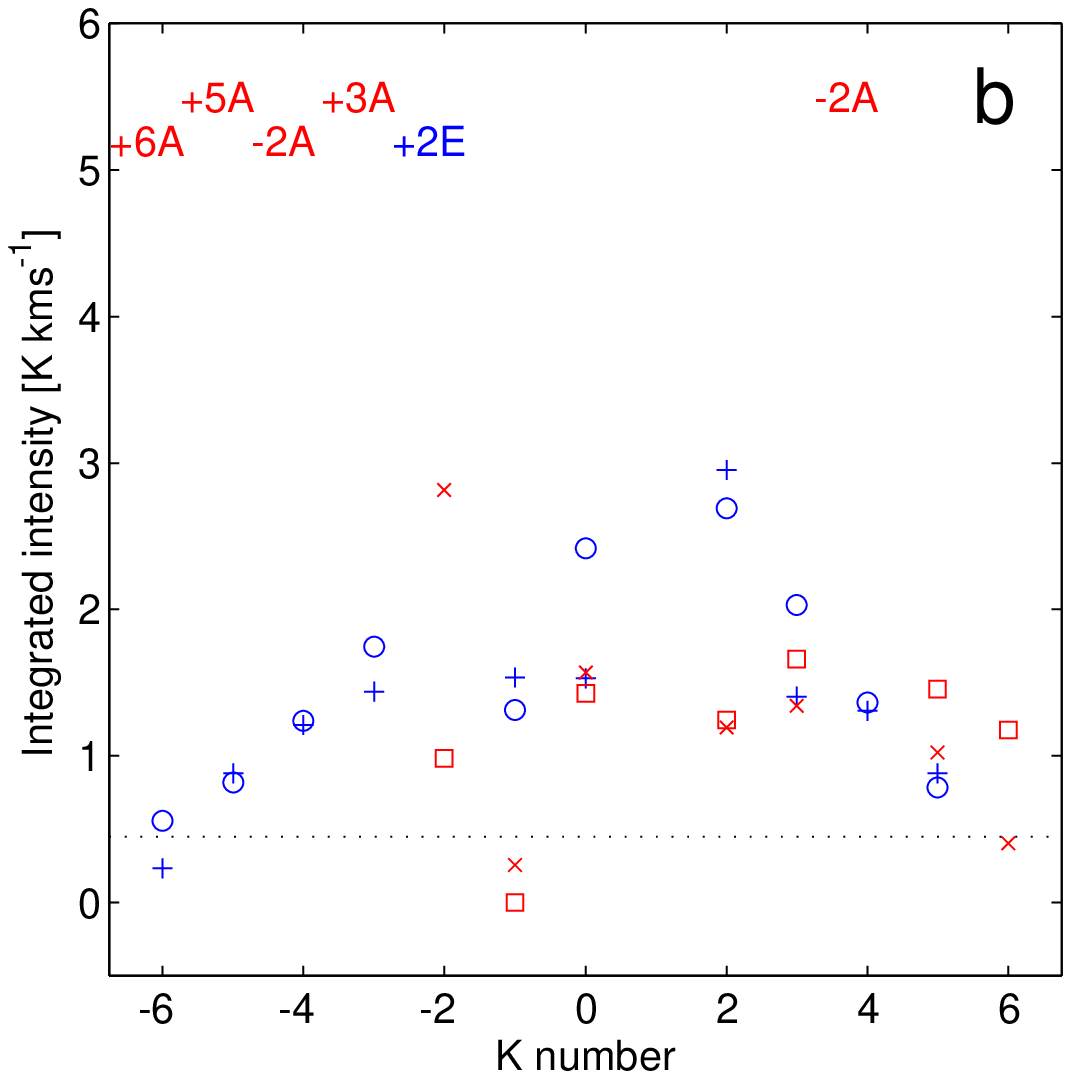}
  \includegraphics[width=8cm]{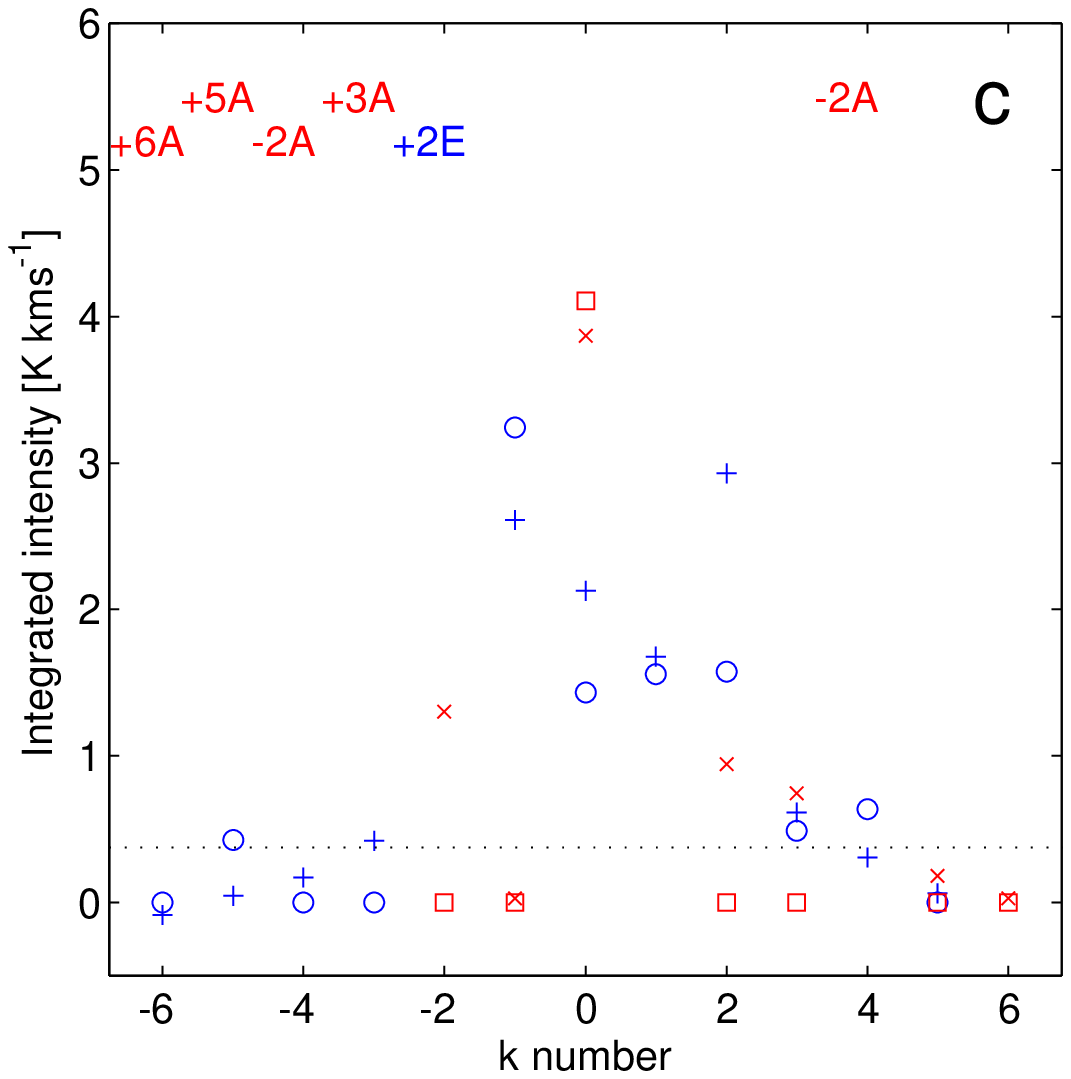}
  \includegraphics[width=8cm]{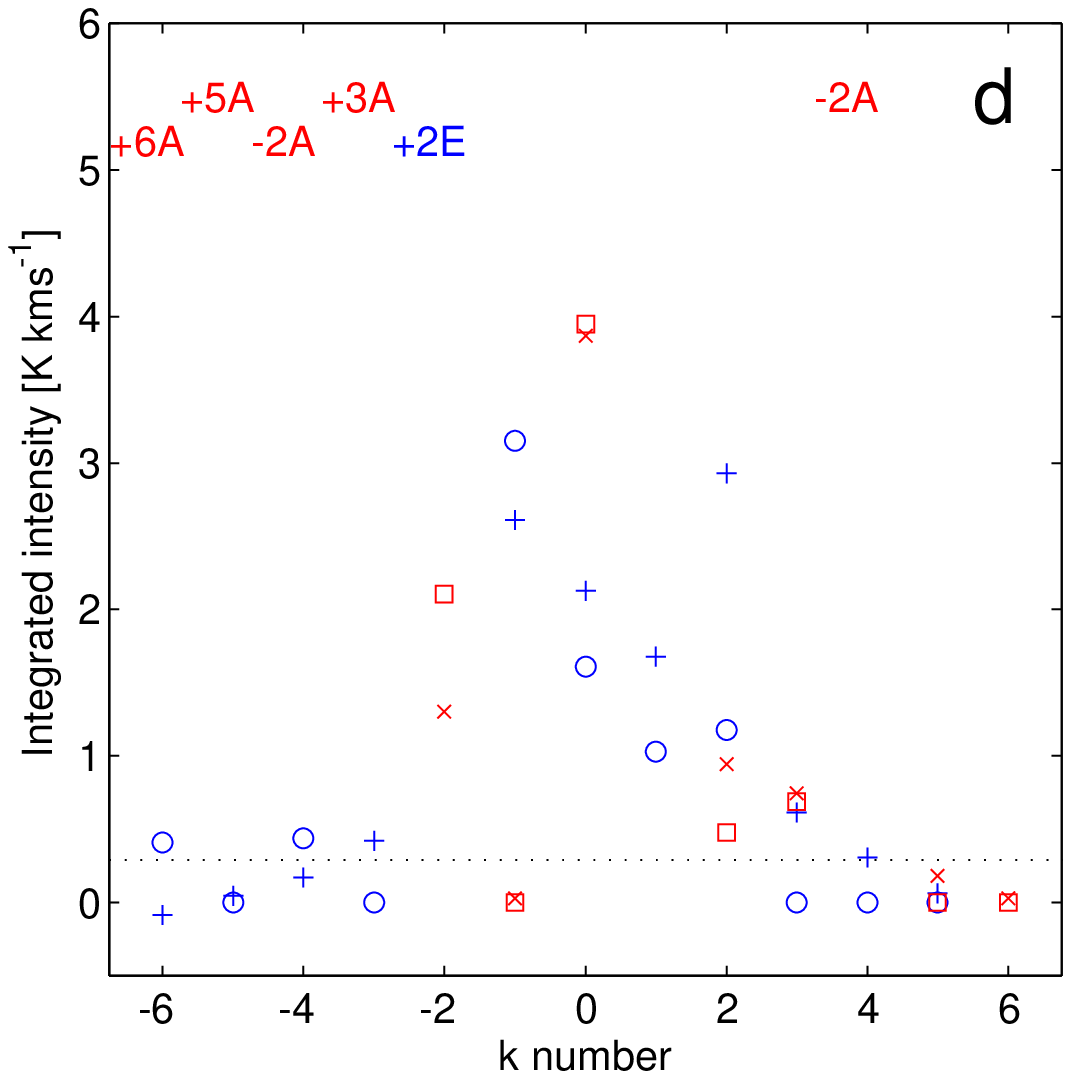}
  \caption{Plots of the integrated intensity for different \methanol\
    $7_\mathrm{K} \to 6_\mathrm{K}$ transitions and the best $\chi^{2}$ model values. In
    the panels the observed flux (open symbols) is plotted, as well as the non-LTE
    model values (plusses and crosses). Also indicated is the 3$\sigma$ detection limit. 
    Circles and plusses (blue) are used for E-type lines and squares and crosses (red) for A-type. Labels
    on the top identify blended lines, for which the combined flux
    and model value are displayed. The
    K=+1 line which is blended with an SO$_2$ was not included in the
    analysis for the centre region components (top panels). 
    The parameters of the fits are listed in Table \ref{radextable}.  
    {\it a:} main, $-10.5$\,\kms\
    velocity component at the ``Centre'' position with beam dilution factor of
    1, {\it b}: red-shifted, $-5.2$\,\kms\ velocity component at ``Centre''
    with a beam dilution factor of 50, {\it c}: the ``NE Outflow'' and
    {\it d}: ``Envelope''.}
  \label{fluxcomp}
\end{figure*}

Next, we use the non-LTE calculations to fit the spectra
for the four regions identified in Sec. \ref{secdistribution} (both
velocity components at ``Centre'', the ``NE outflow'', and the quiescent
gas in the ``Envelope''). In a search for minima of the reduced $\chi^2$
we have used a model grid that includes column densities from
10$^{13}$ to 10$^{19}$~cm$^{-2}$ at every half dec, densities
from $10^4$ up to 10$^{9}$~cm$^{-3}$, kinetic temperatures between $30 - 300$K,
and radiation temperatures in the range $3 - 400$K.
Based on our findings in Sec.~\ref{rotsection}
we have adopted a beam dilution factor of 50 for the 
red-shifted emission in the ``Centre'', corresponding to the 2\arcsec\ extent 
of the region where the masers appear.
It is found that the models are able to
reproduce the lines at all four positions to within 5-10\%
accuracy. In Fig. \ref{fluxcomp} we plot the observed integrated flux
and the synthetic integrated flux calculated with RADEX versus
7$_\mathrm{K}\to$6$_\mathrm{K}$ number. Also indicated is our
3$\sigma$ detection limit.

Like before, this analysis is complicated somewhat by the blending of lines, both by
methanol and in the case of the K=+1 E by blending with SO$_2$. In the
case of the K=+1 E line we have not included it in the analysis of the
centre position components as
the SO$_2$ contributes significantly to it. For the  ``NE outflow'' and
for the ``Envelope'' the SO$_2$ does not appear to contribute significantly, as
the other SO$_2$ line is not detected, and we have included it in the analysis. 
At the position of \object{HW2} the methanol gas is highly excited and
the blending of methanol lines becomes more severe and confusion prevents
one from detecting the
higher (K$\geq 5$) lines. Also, at this position the K=$\pm2$ lines
show a large ($\sim 50$\%) deviation from the RADEX model. The reason
for this discrepancy is yet unclear, but could be another indication
that we are not complete in treating radiative processes.

\begin{table*}
\begin{minipage}[t]{\columnwidth}
\caption{Results of the non-LTE models.}             
\label{radextable}      
\renewcommand{\footnoterule}{}  
\begin{tabular}{l l l l l l l}        
\hline\hline 
Position & $\Delta \alpha$\tabnote{a} & $\Delta \delta$\tabnote{a} & log(N$_{\mathrm{M}}$) & T$_{\mathrm{kin}}$ & log(n$_{\mathrm{H_{2}}}$) & T$_{\mathrm{rad}}$ \\
             & \arcsec & \arcsec & cm$^{-2}$ & K & cm$^{-3}$ & K \\
\hline
Centre, main & +3	& +3 	& ${16.0}$ 	& $90$ 	& ${4.0}$	& 30 \\ 
Centre, red-shifted \tabnote{b}	& +3	& +3 	& ${17.0}$ 	& $130$ & ${7.0}$ 	& 100 \\ 
Outflow (NE) 	& +27	& +9 	& ${15.0}$ 	& $90$	& ${5.0}$ 	& 30 \\ 
Envelope 	& +21	& -3 	& ${15.0}$ 	& $90$	& ${5.0}$ 	& 30 \\  
\hline 
\end{tabular}
\\
\tabnote{a} Offset from centre of map RA 22$^h$56$^m$17.88$^s$, DEC +62\degr01\arcmin49.2\arcsec.\\
\tabnote{b} This component was modeled with a beam dilution factor of 50.
\end{minipage}
\end{table*}

The results of our non-LTE analysis are summarised in Table
\ref{radextable}. The coolest and most diffuse gas is found in the
envelope and outflow, where one should take into account that the
granularity of the model grid probably prevents the distinction
of small temperature and density differences. In comparison, the gas associated with the rotation
temperature peak is much denser and also warm. We find
the highest density and temperature in the ``Centre'' \object{HW2}
position where the maser emission
occurs. However, given the large beam dilution and the limited 
set of available levels for collisional and radiative excitation, 
these numbers should be treated cautiously. 

The non-LTE analysis confirms that there are no large column density
gradients across the source. But although generally the values are in agreement with our rotation
diagram analysis, they differ in details. In particular, the non-LTE
analysis finds local maxima of column density in the ``Centre'' components.
Although some shortcomings are demonstrated for the inner components, the rotation diagram maps are
useful for outlining the qualitative large-scale properties and constraining
the distribution of the excitation; running pixel based non-LTE models
is beyond the scope of this paper.

\subsection{\hmol\ column density}

In order to estimate the methanol abundance we require a measure 
of the \hmol\ column density to combine with the estimates of the methanol column density.  
Therefore we used the SCUBA map of \citet{bottinelli04}
and derived the \hmol\ column distribution using the formula of
\citet{henning00} (eq. 2), where $S_{\nu}$ is the flux in Jy
beam$^{-1}$, $\kappa_d(\lambda)$ the dust opacity at the observed
wavelength, $\Omega_{mb}$ the solid angle of the main beam,
$B_{\nu}(T)$ the Planck function (Black body), $m_H$ the mass of the
hydrogen atom, and $R$ is the gas to dust ratio. In the calculation
we have used: $\kappa_{d}(875\mathrm{\mu m})=1.4$ and $R$=150
\citep{henning00}.

\begin{eqnarray}
N(H_2)&=&\frac{S_\nu}{\kappa_{d}(\lambda) \Omega_{mb} B_{\nu}(T) 2 m_H} R
\end{eqnarray}

The calculations were run for two cases. In the first case, we
assume a constant dust temperature of \Tdust $=30$~K to derive the
\hmol\ column density and the methanol abundance, $X_\mathrm{M}$. In
the second case we have assumed the dust temperature \Tdust\ to be
equal to the rotation temperature that we derived from the rotation
diagram analysis in Sec. \ref{rotsection} and used that as input to
the black body function.

\begin{figure}
   \centering
   \includegraphics[width=8cm]{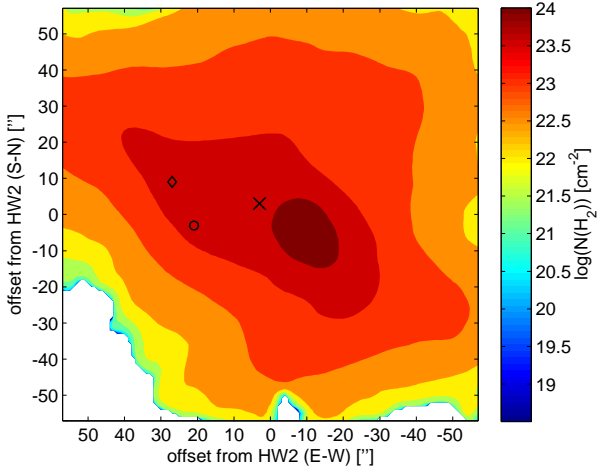}
      \caption{H$_2$ column densities derived from the SCUBA dust continuum for a constant dust temperature of 30~K.}
         \label{h2const}
\end{figure}

\begin{figure}
   \centering
   \includegraphics[width=8cm]{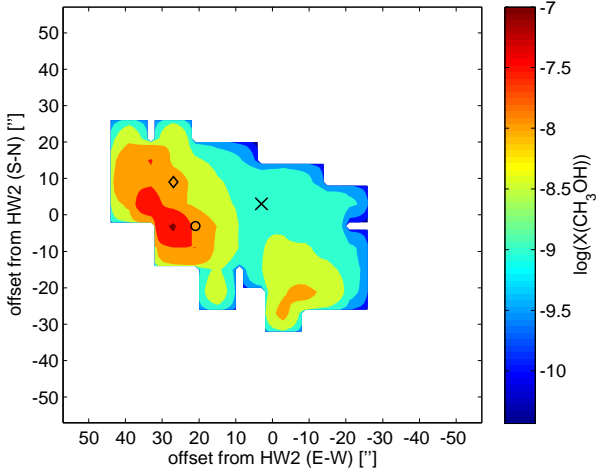}
      \caption{Methanol abundance derived from the methanol column densities obtained in the 
      rotation diagram analysis and the H$_2$ column densities obtained for the constant 
      dust temperature 30~K.}
      \label{xconst}
\end{figure}

\subsection{Methanol abundance}\label{secabundance}

The results of the constant temperature hydrogen column density and
methanol abundance calculations are presented in
Figs. \ref{h2const}-\ref{xconst}. The resulting hydrogen column
density Fig. \ref{h2const}, is the SCUBA map (by \cite{bottinelli04})
scaled by a constant factor to convert it from \mbox{Jy beam$^{-1}$}
to $N$(H$_2$) in cm$^{-2}$. The result is a very smooth map with
\hmol\ column densities between $\sim 10^{21}$~cm$^{-2}$ and $\sim
10^{24}$~cm$^{-2}$. The general morphology of the dust emission is
similar to that of the methanol gas, though notably the dust
emission does not peak at the position of \object{HW2} but further to
the SE at position (-9\arcsec,-4\arcsec).

Similarly, the
methanol abundance map derived with a constant dust temperature,
Fig. \ref{xconst}, shows the same structure as the column density map,
since there are no large variations in the SCUBA dust map. We derive a
methanol abundance of $X_\mathrm{M} \sim$10$^{-8.5}$ for the central
region. As noted before, the distribution is relatively flat, and 
probably not very accurate for the centre position.

In the second case, when we used the methanol rotation temperature as a measure of the dust
temperature, we find the resulting \hmol\ column density to be
anti-correlated to the rotation temperature, as the high rotation
temperatures derived for the centre region then result in a
lower \hmol\ column.  In this way we would derive a \hmol\ column of
$\sim 10^{22}$~cm$^{-2}$ \citep[much lower than inferred earlier $\sim
10^{24}$~cm$^{-2}$;][]{martin-pintado05}. 
Although we have already encountered the limited validity of the 
rotation diagram for the ``Centre'' region in
Sec.~\ref{secnonlteresults}, it is clear that there must be a mismatch between
the gas and dust temperature on the scales of the JCMT maps. Both the
rotation diagram and non-LTE calculations indicate higher temperatures
towards the centre, for which there is little evidence in the dust
images, which even peak at an offset of $\sim$10\arcsec\ from the
rotation temperature maximum. Besides the possibility that the
rotation temperature is not a good estimate of the kinetic temperature
of the gas, it seems plausible that the kinetic temperature of the gas
is not equal to the dust temperature at these locations. Therefore, 
we adopt a constant dust temperature of 30~K to estimate
the H$_2$ column density and \methanol\ abundance, realizing
that this method will overestimate $N$(H$_2$) at the centre of the map.


\section{Discussion}\label{secdiscussion}

\subsection{Outflow morphology}

We have presented maps of the large scale distribution of the
thermal methanol associated with \object{Cep A} East 
(Figs.~\ref{line03plots}, \ref{Trotmap} and \ref{Nrotmap}). The emission is
constrained to a region of size 0.29 $\times$ 0.15~pc (85\arcsec\ $\times$ 43\arcsec),
roughly centred on the location of the radio continuum source
\object{HW2}, the most massive YSO in the region. Moreover, the
velocity field shows that the methanol gas is entrained in a large
scale bipolar outflow. The position angle of this outflow appears
consistent with the \object{HW2} geometry and earlier results
\citep{gomez99}.

While the largest flux of the K=-1 E line Fig. \ref{line03plots}a is
associated with the receding part of the outflow (towards the SW),
coincident with the SCUBA peak, the most energetic gas appears at the
opposite side of \object{HW2}. This is evident already in the velocity
width distribution of the brightest line, that has a line width of
$>$4~\kms, to the NE of \object{HW2}, Fig.~\ref{line03plots}c. It
becomes more clear when the temperatures are derived from rotation
diagram analysis, the rotation temperature peaks to the NE of
\object{HW2} at $\sim$200~K (Fig. \ref{Trotmap}).

Several authors \citep{patel05,jimenez-serra07,torrelles07} have
inferred obscuring gas in a flattened circumstellar structure of
$\sim$1\arcsec\ (700~AU) surrounding \object{HW2}. The gas shows signs
of rotation and is perpendicular to the high velocity outflow
($\sim$500~\kms) seen in the radio continuum \citep{curiel06}. The
molecular outflow that we observe in methanol is blue-shifted to the NE,
which implies that this side of the outflow is pointing towards the observer,
Fig. \ref{line03plots}b. \citet{torstensson11} propose a model in
which the methanol maser emission arises in a ring-structure in the 
equatorial plane of \object{HW2}. In contrast to the circumstellar 
molecular material, the masers do not show signs of rotation; rather 
the observations seem to indicate an infall signature. It is argued
that the masers occur in the shock interface between the circumstellar
disk and inflow regulated by a large scale magnetic field \citep{vlemmings10}.
\citet{torstensson11} derive an inclination of 67.5\degr
and a position angle of 9.3\degr\ for the ring. The masering gas occurs
on size scales of $\sim$2\arcsec\ and is very likely clumpy on smaller
scales $\sim$0.1\arcsec \citep{minier02}. At the resolution of the
observations presented in this paper the masering gas can clearly not be
resolved, nevertheless the geometry that is observed on smaller scales
suggest that to the NE of \object{HW2} we have a line of sight into
the outflow probing highly excited gas close to the protostar, which
explains why we see such a high rotation temperature in this region.
Evidence for such wide-angle molecular outflow has recently been
presented by \citet{torrelles10}.

\subsection{Methanol distribution}

The highest methanol temperatures appear to occur close
to the HW2 central source, but we have demonstrated in Sec. \ref{secabundance}
that the dust emission is not showing the same distribution.
It seems plausible that the
highest methanol rotation temperatures occur in a region where the
dust is at least partly destroyed, possibly at the onset of the
outflow where the evaporation from the grains occurs that produces the methanol gas in
the first place. 
The considerations on the dust distribution force us to adopt a
constant dust temperature in our abundance analysis. On small radii
this may have two effects: firstly, the gas to dust ratio may be
higher due to grain destruction, which increases $N$(H$_2$) and lowers
the abundance. Secondly, the dust temperature may be higher than 30~K,
which decreases $N$(H$ _2$) and increases the abundance. Increasing
the gas-to-dust ratio from 150 to 200/250 decreases the abundance with
a factor of 0.75/0.60 respectively. Increasing the dust temperature to
60/100/300~K increases the resulting abundance by a factor of 2.3/4.1/13.0
respectively. High optical depth of the dust will lower the
derived abundance. These arguments show that at the peak rotation
temperature, the methanol abundance is quite sensitive to
increased dust temperatures, but is probably correct to within an
order of magnitude, without taking beam-dilution effects into account.

The derived abundances of $\sim10^{-8.5}$ are higher than expected from
gas-phase chemistry, but lower than seen in the solid state
\citep{gibb04}. High optical depth of the lines is ruled out by the
excitation calculations, except in the central region. However, it is possible that there the
local methanol abundance is still higher. For example,
the methanol maser occurs on much smaller
scales than those we are probing in these JCMT observations. The total
extent of the \methanol\ maser is $\sim<$2\arcsec. Moreover, we found that the same
small scale enhancement can explain the peculiar excitation of the ``Centre''
region. This strengthens the idea that the
highly excited methanol gas emission is coming from the same region. 
Assuming that the dust and the corresponding \hmol\ is smoothly 
distributed, this component can have higher column density
because of the optical depth and beam-dilution, resulting in 
methanol abundance estimate of $\sim10^{-7}$ locally. This
value is similar to previous values by other authors for maser regions
\citep{menten86,menten88}
and still higher than expected from gas-phase chemistry and lower than
seen in the solid state. Possibly the methanol evaporated from grain
mantles some time ago ($\sim$1000~yr) and is now destroyed by chemical
reactions and/or photo-dissociated.

As noted earlier, all the methanol gas seem to be entrained in an
outflow, spanning 0.29~pc. The outflow is
oriented close to the plane of the sky, assuming an inclination of
67.5\degr\ \citep{torstensson11} we find an
outflow velocity of $\sim$5.8 \kms. The
result implies a dynamical timescale of $\sim 2.4 \times 10^{4}$
yrs. The dynamical timescale agrees well with the chemical timescale
(a few $10^4$ years) for which such a methanol enhancement is believed
to be possible \citep{vandertak00c}. This supports the argument of a
common driving source for all the methanol gas in the region.
Alternatively, the methanol may be released in the wake/outflow cavity
of the jets described by \citet{cunningham09}: the uv-radiation
produced in the shock interface may shine back into the cavity and
photo-desorb the methanol, causing a local enhancement in the methanol
column density.

\subsection{Physical conditions}

The rotation diagram analysis shows rather warm gas entrained in
a large scale outflow. Although the non-LTE analysis demonstrates
that the methanol band used in this study is quite sensitive to 
collisional excitation, even in the low temperature regime, it 
confirms qualitatively the large scale presence of warm methanol gas.
Over large areas of the source the total methanol column density 
appears to be fairly constant at $\sim 10^{15}$~cm$^{-2}$, 
assuming a uniform distribution with unity beam-filling factor.

The fidelity of the rotation diagrams is breaking down particularly in the
centre area, where higher excitation lines appear. Especially the 
red-shifted component requires non-LTE excitation and/or beam-dilution
effects to be explained. At these locations a complete analysis requires 
high resolution data and non-LTE modeling including
excitation by infrared radiation. That more methanol levels should be taken
into account is
demonstrated by the detection of the torsionally excited 
$-1$ A v$_\mathrm{t}$=1 line at 337.969 GHz in the red-shifted
component \citep{leurini07}. Radiative pumping is also expected from models of methanol maser excitation
\citep{cragg05, sobolev94, sobolev97}

\subsection{Origin of maser emission}

Although a larger area around Cep~A has been searched 
for maser emission \citep{torstensson11}, only one place in this source,
so abundant in methanol, 
does show maser emission. We find the maser position to contain warmer gas
than the other positions,
probably the highest column density, and certainly the density
appears to have a maximum here at $\sim 10^7$~cm$^{-3}$.
Moreover, this position does stand out as the position where
a second kinematic component at $-5.2$\,\kms\ is found. Special properties for
this component are the detection of the torsionally excited line,
and the requirement to invoke a large beam dilution to explain the 
ratios of the higher excited lines. This dilution factor matches
quite well with the size of the area of $<2$\arcsec\ over 
which masers are detected \citep{sugiyama08,
  torstensson11}. Combining these facts, we
identify this component with the origin of the maser 
action. We note that the densities we derive for this component
are still below $10^{9}$~cm$^{-3}$, where the maser lines would
be quenched according to model calculations \citep{cragg05}.
The models have methanol
temperature ranges and
methanol column densities that are consistent with our estimates, but
it is hard to test the dust temperature, which must be
$>$100~K for the maser excitation. We note that many of the 
values estimated here are inferred indirectly
and the relevant processes may occur on even smaller scales
than reflected by the beam-dilution factor.

The  central abundance of $X_{\mathrm{M}} \sim
10^{-7}$ is not in excess to that found in low-mass
star-forming regions, but the methanol column density 
$N_{\mathrm{M}} \sim 10^{17}$~cm$^{-2}$ clearly is \citep{maret05}.
This could point to the condition that the high-density, high-temperature region in low-mass star-forming
regions is too small to produce a path length long enough, with
sufficient methanol column density, to produce maser
emission. \citet{jorgensen02} show the high-density, high-temperature
region for a low-mass protostar to be $\sim$10-30~AU. In contrast, for
a high-mass protostar, the region extends over $\sim$1000~AU
\citep{doty02}, possibly facilitating the necessary amplification. 
Alternatively, \citet{pandian08} suggest that 
in low-mass stars the critical infrared pumping is only available in
the inner region, where the density is so high that the maser will
be quenched. Both effects favour maser emission in the environments of high mass protostars
and explain that the masers occur on rather large distances from the
central object \citep{torstensson11}.

To further constrain the physical conditions at the position of the
methanol maser and discern radiation and density effects, high-J and 
$\nu_t = 1$ line observations are required, for
example with the CHAMP+ instrument on the APEX telescope. To probe 
the excitation of methanol gas on small scales (comparable to 
that of the maser emitting region) high resolution interferometry 
observations are needed, with instruments such as the SMA and the 
IRAM interferometers, and in the longer term, for lower 
declination sources, ALMA. The interpretation of these observations 
requires models that include the excitation of CH$_3$OH by infrared radiation.

\begin{acknowledgements}
	The authors wish to thank an anonymous referee for comments that
	led to substantial improvements of the analysis methods and the 
	robustness of the conclusions of this paper.
  This research was supported by the EU Framework 6 Marie Curie Early
  Stage Training programme under contract number MEST-CT-2005-19669
  ``ESTRELA''. WV acknowledges support by the {\it Deutsche
    Forschungsgemeinschaft} through the Emmy Noether Research grant VL
  61/3-1.
\end{acknowledgements}

\bibliographystyle{aa}
\bibliography{kalle}

\end{document}